\documentclass{openjournal}
\usepackage[utf8]{inputenc}
\shortauthors{Wang et al.}

\usepackage{graphicx}
\usepackage{hyperref}
\usepackage{amsmath}
\usepackage{color}
\usepackage[caption=false]{subfig}
\usepackage{comment}
\usepackage{verbatim}

\begin{document}

\title{Neural Network Based Point Spread Function Deconvolution for Astronomical Applications}

\author{Hong Wang$^{1}$}
\author{Sreevarsha Sreejith$^{2}$}
\author{Yuewei Lin$^{1}$}
\author{Nesar Ramachandra$^{3, 4}$}
\author{An\v{z}e Slosar$^{2}$}
\author{Shinjae Yoo$^{1}$}

\affiliation{$^{1}$ Computational Science Initiative, Brookhaven National Laboratory, Upton, NY 11973}

\affiliation{$^{2}$ Physics Department, Brookhaven National Laboratory, Upton, NY 11973}
\affiliation{$^{3}$ Computational Science Division, Argonne National Laboratory, Lemont, IL, USA}
\affiliation {$^{4}$ High Energy Physics Division, Argonne National Laboratory,Lemont, IL, USA}

\begin{abstract}
Optical astronomical images are strongly affected by the point spread function (PSF) of the optical system and the atmosphere (seeing) which blurs the observed image. The amount of blurring depends both on the observed band, and on the atmospheric conditions during observation. A typical astronomical image will likely have a unique PSF, that is non-circular and different in different bands. At the same time, observations of known stars also give us an accurate determination of this PSF. Therefore, any serious candidate for production analysis of astronomical images must take the known PSF into account during the image analysis. So far, the majority of applications of neural networks (NN) to astronomical image analysis have ignored this problem by assuming a fixed PSF in training and validation.  We present a neural-network based deconvolution algorithm based on Deep Wiener Deconvolution Network (DWDN). This algorithm belongs to a class of \emph{non-blind} deconvolution algorithms, since it assumes the  PSF shape is known.   We study the performance of different versions of this algorithm under realistic observational conditions in terms of the recovery of the most relevant astronomical quantities such as colors, ellipticities and orientations. We investigate custom loss functions that optimize the recovery of astronomical quantities with mixed results.
\end{abstract}

\section{Introduction}

 The advent of large telescopes and big data sets are bringing about a transformative era for astronomical survey science. New datasets obtained from upcoming ground-based and space-based observing facilities will extract data from large volumes of the observable sky at unprecedented depths and cadences. An important addition to the large array of data observed using ground-based observatories will be the Legacy Survey of Space and Time (LSST) from the Vera Rubin Observatory \citep{lsstbook2009}, adding to existing datasets from the Dark Energy Survey (DES, \citealt{des2018}) and Hyper-Suprime Cam (HSC,\citealt{hsc2019}). We also expect upcoming space-based surveys such as Euclid \citep{euclid2010} and Nancy Grace Roman telescope \citep{roman2012} to add on to this data influx and produce excellently uniform datasets. While the resolution of space-based telescopes is usually diffraction limited, large aperture ground based survey telescopes are limited by seeing. Seeing is a distortion of the image caused by the perturbation of an optical wavefront as it passes through the turbulent atmosphere and results in a finite point-spread function (PSF). Due to the large field of view, these surveys will not be able to employ adaptive optics resulting in an arcsecond sized PSF. 

The two main features of the PSF is that it is variable, because the atmosphere above the telescope is variable. It varies both in size, shape and orientation.  It is typically asymmetric at the level that matters for analysis and varies in size and asymmetry from one pass band to another. On the other hand, it is also known in each exposure, because invariably every exposure contains stars that are excellent point sources and therefore cane be used to directly probe the PSF.

In astronomical image analysis we want to recover the unbiased estimates of the intrinsic parameters of objects, such as their shapes, positions and fluxes in the presence of noise and varying PSF. Machine learning techniques hold promise for speeding up these processes. Several machine learning based methods have been proposed recently that use neural networks for astronomical image analysis. These mostly grapple with the galaxy deblending problem. ~\cite{reiman2019deblending} designed a branched deblender with generative adversarial networks, which can deblend images with two overlapped galaxies. ~\cite{boucaud2020photometry} developed a framework for measuring the photometry of blended galaxies as well as perform segmentation with a standard convolutional neural network (CNN) and a U-Net. ~\cite{2005.12039} introduced an algorithm where a Variational Auto Encoder(VAE)-like neural network was used for galaxy deblending. We ourselves, have proposed an approach using residual dense neural networks \citep{wang2021}. 

Most of the current generation neural-network based galaxy deblenders assume a number of simplifying assumptions, including a constant PSF. This problem is usually ``swept under the rug'' by assuming that several training sets can be provided, one for each PSF. However, this is likely impractical for reasons we will cover in the following sections. It is also clear that simply ignoring this problem will lead to biased and sub-optimally inferred parameters for individual objects. Therefore the proper treatment of the PSF is one of the few issues that need to be addressed before neural network based approaches can be translated from the realm of toy problems to realistic approaches. Other issues include masking and various detector artifacts. In this work we specifically address the question of PSFs, and build a network that takes a noisy, convolved image of an astronomical object, uses PSF information to deconvolve it and produce a true image of the galaxy at the resolution supported by the PSF. 

 Note that there is an intrinsic connection between the issues of deconvolution and blending. While it may be apparent that two objects can merge into a single entity if the point spread function (PSF) is too large, there are also more subtle effects that can occur. For example, at a fixed PSF shape, the likelihood of two distinct objects blending depends on whether the line that connects them is parallel or perpendicular to the elongation of the PSF.

\section{Description of the problem}

We use a simple linear model for the observed astronomical image, a convolution with a PSF followed by the addition of observational noise:
\newcommand{\PSF}{{\rm PSF}}
\newcommand{\WD}{{\rm WD}}
\begin{equation}
    I_{i} = I_{gt,i} \circledast \PSF_i + N_i, \label{eq:problem}
\end{equation}
where $\PSF_i$ is the PSF for the channel $i$, $N_i$ is the noise and index $gt$ refers to the ground truth.
In this work we assume that while noise need not be of the same level in each channel, it is white and normally distributed. One of the main drawbacks in this kind of approaches is the assumption that the noise independent of the signal. This is true when the dominant source of the sky are not photons from the object itself, i.e.,  for sky-noise and read-out noise dominated images, which is the case for ground-based observations. However, this assumption doesn't hold for space-based imaging and therefore more work will need to be done to establish the method in that case.

From now on, we will drop the band index $i$, denoting that each operation is performed independently on each band.  It is important to note that both the PSF and the amplitude of noise are band-dependent. 

This model is not strictly true because the detector effects such as brighter-fatter, amplifier non-linearities, etc. make the transformation weakly non-linear, but these effects are small. Moreover, the shape of the PSF varies across the focal plane: in this work we assume that the region of interest is always small enough that the PSF can be assumed as constant.

Since convolution is a linear operation, it is formally invertible. However, this process is not stable. Denoting the quantities in Fourier space with a superscript $X^F(k) = \mathcal{FT}(X(x))$, the  Equation~(\ref{eq:problem}) becomes
\begin{equation}
    I^F(k) = I^F_{gt}(k) \cdot \PSF^F(k) + N^F(k), \label{eq:fourier}
\end{equation}
in Fourier space since the convolution maps to a multiplication.

In this space, the naive inversion is therefore given by 
\begin{equation}
    I_{\rm deconvolved}^F = (\PSF^F)^{-1}(k) I^F(k) = I^F_{gt}(k) +  (\PSF^F)^{-1}(k){N^F(k)}
    \label{eq:fourierD}.
\end{equation}
Since the noise is added \emph{after} the convolution with the PSF, the recovery of spatially high-frequency modes that were blurred by the PSF are replaced with a blown-up noise that dominates the output image. In other words, because $N^F(k)$ is approximately constant, while $PSF^F(k)$ drops with k, the spatially high-frequency modes in the original image are truly lost information and therefore cannot be recovered. The deconvolution thus needs to be regularized. 

One of the classical regularization techniques is Wiener deconvolution (\citealt{Wiener49}). In this technique, we deconvolve the modes which have high signal-to-noise and suppress those that are noise dominated.  It is a special limit of the Wiener filter which  minimizes the mean square error residuals between the estimate and the truth for stationary random processes. In Fourier space, the deconvolution operator can be written as (dropping the band subscripts for clarity):
\begin{equation}
    I_{\rm deconvolved}^F = \left( \frac{P_{s} (k)}{P_{s}(k)+P_{n}(k) |\PSF^F(k)|^{-2}} (\PSF^F)^{-1}(k) \right) I^F,
    \label{eq:wiener1}
\end{equation}
where $P_s(k)$ and $P_n(k)$ are spectral densities of signal and noise respectively.

The Wiener filter is optimal for images which are stationary random processes and has the advantage of being linear in the input pixel values. The Deep Wiener Deconvolution Network is a neural network extension of Wiener Deconvolution  which we describe in the following section.

\section{Band-wise Deep Wiener Deconvolution Network}\label{sec:method}
The network utilized in this paper is an extension of the Deep Wiener Deconvolution Network (DWDN, \citealt{dong2020deep}). DWDN provides a brand new solution for the non-blind image deblurring problems. Instead of denoising and deconvolving the image in the image space like the existing methods do, DWDN applies Wiener Deconvolution explicitly in the feature space. As a simple but effective integration for classical Wiener Deconvolution and deep learning, DWDN achieves outstanding performance with fewer artifacts in solving the non-blind image deblurring problem. It contains two components, the feature-based Wiener deconvolution module and a multi-scale feature refinement module, where the former carries out the deconvolution process and the latter restores high-quality images using features from the previous module. The procedure is given by

\begin{equation} \label{eq:DWDN}
    \begin{gathered}
    h^l = \mathcal{FT}^{-1}\left(\frac{\overline{\PSF^F(k)}}{\overline{\PSF^F(k)}\PSF(^Fk)+\frac{s^n_l}{s^x_l}}\right) \\
    \hat{I} = \mathcal{R}(h^1, \dots, h^L) 
    \end{gathered}
\end{equation}

where $h^l$ is the deconvolved feature for $l$-th channel and $l \in [1,\dots, L]$. $L$ denotes the number of channels of the feature. $\mathcal{F}$ refers to discrete Fourier transformation and $\overline{\PSF(k)}$ is the complex conjugate of $\PSF(k)$. $s^n_l$ and $s^x_l$ are the variance and standard deviation of $l$-th channel of the features in latent space for noise and input image separately which can be estimated by blurred features and mean-filtered features. $\mathcal{R}$ refers to the feature refinement module and $\hat{I}$ denotes the output of the model.

\begin{figure}[htbp]
  \centering
  \includegraphics[width=0.8\linewidth]{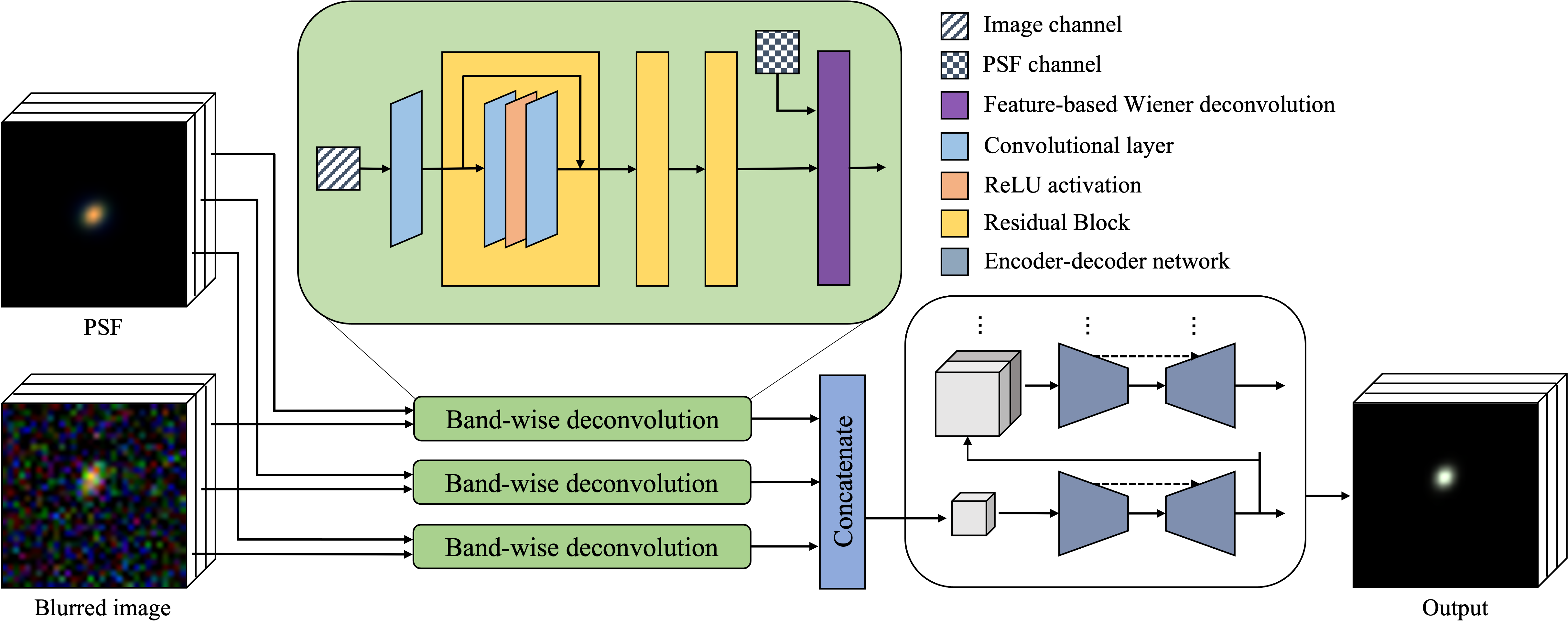}
  \caption{The framework of the model used in this work. The three channel-wise deconvolution modules share the same architecture. The feature refinement module contains convolution-based encoder-decoder network for multiple levels. The kernel size is $5\times5$ for all convolutional layers.}
  \label{fig:framework}
\end{figure}

A restriction that limits the direct use of DWDN on astronomical images is that DWDN was originally designed for achromatic motion blurring problems.
For galaxy images, different bands have different PSFs. However, in DWDN, the feature maps extracted from the input have already mixed up and no longer have distinguishable bands as the image does. Therefore, we propose an enhanced variant of DWDN that can work on band-dependent PSFs. Figure \ref{fig:framework} shows the architecture of our model. In order to apply feature-based deconvolution on different input bands separately, we parallel three deconvolution modules. The number of the deconvolution modules varies based on the number of bands in the input image and PSF. Each band of the input image with the corresponding PSF band will go through one of those modules to complete the feature-based deconvolution. Eq.~\ref{eq:band-DWDN1} expresses the deconvolution for the  $l$-th feature channel in the $i$-th deconvolution module
\begin{equation} \label{eq:band-DWDN1}
    h_i^l = \mathcal{FT}^{-1}\left(\frac{\overline{\PSF^F_i(k)}}{\overline{\PSF^F_i(k)}\PSF^F_i(k)+\frac{s^n_{i,l}}{s^x_{i,l}}}\right)
\end{equation}
\begin{equation} \label{eq:band-DWDN1_2} 
    H_i = \mathcal{C}[h_i^1, \dots, h_i^L].
\end{equation}

The features will then be concatenated as Eq.~\ref{eq:band-DWDN1_2}, which means that $H_i$ is the output from the $i$-th deconvolution module, namely, deconvolved features for the $i$-th band of the input image and PSF. The deconvolved features will then be passed into the second module as shown in Eq.~\ref{eq:band-DWDN2} with $C$ denoting the total number of deconvolution modules. We keep the refining module $\mathcal{R}$ the same except that the number of channels of its input feature is $C$ times that of the original DWDN because of the concatenation. In this way, the different bands of the input image will execute their own deep Wiener deconvolution and the deconvolved features are utilized to restore the clean images.

\begin{equation} \label{eq:band-DWDN2}
    \hat{I} = \mathcal{R}(H_1, \dots, H_C).
\end{equation}

Specifically, there are $C=3$ parallel deconvolution modules. Each module is composed of $m$ residual blocks following a convolution layer and each residual block contains $2$ convolutional layers. The multi-scale feature refinement module starts from a group of coarse-to-fine pyramid-like features which are downsampled from the concatenated feature using bicubic downsampling. The features will go through an encoder-decoder network with skipping connections to restore the clean image. In the refinement module, the output from the previous level will also be upsampled and utilized for the computation of the next level. The output of the final level will have the same shape as the original input image and is the final prediction from the model.

\section{Custom loss function}
\label{sec:custom_loss}
In majority of image analysis problems, the networks are trained with one of the standard loss functions that encode some intuitive notion of similarity between the true and reconstructed images. Typical examples include $\ell_1$ (Least Absolute Deviation, LAD, \citealt{L1ref}), $\ell_2$ (Least Square Errors, LSE, \citealt{L21ref,L22ref}) loss functions\footnote{L1 \& L2 functions are also used for regularization, in order to prevent overfitting.}, and cross-entropy functions (binary, categorical, sparse categorical etc., \citealt{CEs})\footnote{https://keras.io/}.  

However, in astronomical image analysis we are interested in certain specific properties of the output image. The first few image moments detailed in the next section describe the astronomical quantities of interest: object brightness, position and shape. It is therefore natural to include them into the loss function for the errors to be minimized during training -- in this manner we train the neural network to focus explicitly on the quantities of interest. This `non-standard' training procedure is described in the following section.

\subsection{Quantities of interest} \label{sec:list}

Let $I_{ij}$ be the flux value in pixel $(i,j)$. We define raw moments as

\begin{equation}
    M_{pq}(I)=\sum_{x=1}^{W}\sum_{y=1}^{H} x^py^q I_{x,y} \label{eq:mpq}
\end{equation}
The following quantities are of interest:

\begin{equation}
   \begin{array}{ll}
    M_{00} & \mbox{Total flux} \\ 
    \left<x\right> = \frac{M_{10}}{M_{00}} & \mbox{astrometric position } x \\
    \left<y\right> = \frac{M_{01}}{M_{00}} & \mbox{astrometric position } y\\
     \mu_{20} = \frac{M_{20}}{M_{00}} - \left<x\right>^2 & \mbox{second central moment } x^2\\
     \mu_{11} = \frac{M_{11}}{M_{00}} - \left<x\right>\left<y\right> & \mbox{second central moment } xy\\
     \mu_{02} = \frac{M_{02}}{M_{00}} - \left<y\right>^2 & \mbox{second central moment } y^2\\
     e_1 = \frac{\mu_{20}-\mu_{02}}{\mu_{20}+\mu_{02}} & \mbox{$+$ ellipticity component} \\
     e_2 = \frac{2\mu_{11}}{\mu_{20}+\mu_{02}} & \mbox{$\times$ ellipticity component} \\
     s = \mu_{20}+\mu_{02} & \mbox{object size} 
    \end{array}  \label{eq:moments}
\end{equation}
These quantities can be trivially calculated from a noiseless image in a way that enables their usage in loss functions. However, we have found that implementing them directly leads to very unstable training (as discussed in the next section). This is likely due to the diverging derivatives for pixels away from the object center, where noise features can have outsized effects.  We also tested the implementation of adaptive moments (\citealt{2003MNRAS.343..459H}), which find a weighting function that optimally weights the object in an iterative manner, suppressing information from the pixels that are far from the object center. We follow \cite{1609.07937} and implement the scheme using a Gaussian weighting function $f(\rho)$ defined as
\begin{align}
& M = \frac{T}{2}\left[
\begin{matrix}
    1+e_1 & e_2 \\
    e_2 & 1-e1\\
\end{matrix}
\right]\\
& \rho = \left[ \begin{matrix}
x-x_0\\
y-y_0    
\end{matrix}\right] \left( M \right)^{-1}
\left[ \begin{matrix}
x-x_0\\
y-y_0    
\end{matrix}\right] \\
& f(\rho) = e^{-\frac{\rho}{2}}\\
& f'(\rho) = -\frac{1}{2}f(\rho),
\end{align}
where $x_0$ and $y_0$ are the object's astrometric center, $\left<x\right>$ and $\left<y\right>$. With the weighting function thus defined, the adaptive moments' iteration is given by 
\begin{align}
    & w \leftarrow f(\rho) \\
    & w' \leftarrow f'(\rho)  \\
    & A \leftarrow \frac{\sum I w}{\sum w^2} \\
    & \mathbf{x} \leftarrow \frac{\sum \mathbf{x} I w'}{\sum I w'} \\
    & M \leftarrow 4 \frac{\sum 
    (\mathbf{x}-\mathbf{x_0})(\mathbf{x}-\mathbf{x_0})^T I w' }{\sum I w},
\end{align}
where $\mathbf{x}$ denotes the vector $(x,y)$.
Usually, this process is iterated until the desired convergence in recovered values of the moments is achieved. In our case we fix the number of iterations to four, which then turns  the adaptive moments into an auto-differentiable procedure. We have checked that the adaptive moments recover the naive moments for isolated Gaussian objects.

\subsection{Training procedure and loss function}

The model is trained in an end-to-end manner and it involves two phases. In the first phase, the model is pre-trained to acquire the ability to deblur and produce a reasonable prediction as the output, while in the second phase, the model is fine-tuned with a more complex loss including the quantities listed in Section \ref{sec:list} for a more accurate description of the galaxy.

The loss function for the first phase is formulated in Eq. \ref{eq:l_p1}. The first part of the loss is $\ell_1$-norm between the output image and the ground truth as shown in Eq. \ref{eq:l1}. We also include the difference of the total flux in this phase written as Eq. \ref{eq:l_m00} as the image properties described are close. In Eq. \ref{eq:l1}, $f$ is the network, $I$ and $I_{gt}$ refer to the blurred input and the ground truth separately. 
$\lambda_0$ controls the trade-off between these two parts.

\begin{equation}
    \mathcal{L}_{\ell_1} = \frac{1}{WHC}\sum_{c=1}^{C}\sum_{x=1}^{W}\sum_{y=1}^{H} \left|f(I)_{x,y,c}-(I_{gt})_{x,y,c} \right| \label{eq:l1}
\end{equation}

\begin{equation}
    \mathcal{L}_{M_{00}}=\frac{1}{C}\sum_{c=1}^{C} \left|M_{00}(f(I))-M_{00}(I_{gt}) \right| \label{eq:l_m00}
\end{equation}

\begin{equation}
    \mathcal{L}_1(I, I_{gt}) = \mathcal{L}_{\ell_1}+\lambda_0\mathcal{L}_{M_{00}} \label{eq:l_p1}
\end{equation}

In the second phase, in addition to $\mathcal{L}_{\ell_1}$ and $\mathcal{L}_{M_{00}}$, the astrometric position $\left<x\right>$ and $\left<y\right>$ \& the second-order central moments $\mu_{11}$, $\mu_{20}$, $\mu_{02}$ are utilized to fine-tune the model. The quantities are defined in Section \ref{sec:list} and the loss for these quantities has the same form as Eq. \ref{eq:l_m00}. Thus, the total loss for the second phase can be written as Eq. \ref{eq:l_p2} where $\lambda$'s are the weights for the different terms.

\begin{equation}
    \mathcal{L}_2(I, I_{gt}) = \mathcal{L}_{\ell_1}+\lambda_0\mathcal{L}_{M_{00}}+\lambda_1\mathcal{L}_{\left<x\right>}+\lambda_2\mathcal{L}_{\left<y\right>}+\lambda_3\mathcal{L}_{\mu_{11}}+\lambda_4\mathcal{L}_{\mu_{20}}+\lambda_5\mathcal{L}_{\mu_{02}} \label{eq:l_p2}
\end{equation}

Values of the $\lambda_i$ parameters are manually adjusted so that those terms are comparable with the dominant $\ell_1$ loss. Using the naive moments, we have found that this leads to very unstable training, with loss function catastrophically increasing. This could be cured by lowering the values of $\lambda$s, at which point this addition to the training had very limited effect on the final result.  We have found that using adaptive moments, the stability of training significiantly increased and allow us to increase $\lambda$ to the level where contribution to the loss function from adaptive moments is comparable in size to the $\ell_1$ loss.

\section{Training and testing data}
The dataset used in this work is simulated using the deep generative models put forth in \cite{lanusse2021}  that creates galaxy images with realistic morphologies\footnote{https://github.com/mcwilliamscenter/galsim\_hub}. The model consists of a hybrid variational autoencoder (\citealt{kingmawelling2013})  with the aggregate posterior distribution modelled by a  latent-space normalizing flow, dubbed as \texttt{Flow-VAE}, and was trained on a dataset based on the  HST/ACS COSMOS  survey (\citealt{Koekemoer_2007,Scoville_2007a,Scoville_2007b}) generated with GalSim (\citealt{galsim2015}).  For more details about the generative model, please see \citealt{lanusse2021} and references therein. 

 The three physical parameters required to render the galaxy images using this code are half-light radius (\texttt{flux\_radius}, denoting the size), magnitude (\texttt{mag\_auto}) and photometric redshift (\texttt{zphot}). The values for these parameters are drawn from respective uniform distributions within the intervals :  $5 \leq \mbox{flux\_radius} \leq 15$, $5 \leq \mbox{mag\_auto} \leq 25$ and $0 \leq \mbox{zphot} \leq 2$ . We offset the galaxies thus generated, from the centre of the postage stamp with a randomly chosen value between $(-5,5)$ in both $x$ \& $y$ directions.  These images are then convolved with a small \texttt{GalSim}  rendered Gaussian PSF with a full width half maximum (FWHM) of $0.4$. The purpose of this initial convolution is to avoid aliasing and to remove modes that no deconvolution process can recover. We use these images as the ``ground truth''.

The deep generative model only provides a single band (band 1) and since we are interested multi-band deconvolutions, two more bands are generated with simple non-linear transformations of the original image of the type
\begin{equation}
    f \rightarrow f \left(1 + \alpha \left( \frac{f}{f_{\rm max}}\right)^a+ \beta \left(\frac{f}{f_{\rm max}}\right)^b \right).     
\end{equation}

Values of $\alpha$ and $\beta$ were chosen arbitrarily with $0 \leq \alpha \leq 0.2$ \& $0 \leq \beta \leq 1$ for band 2 and  $0 \leq \alpha \leq 0.3$ \& $0 \leq \beta \leq 0.5$ for band 3. The values for $a$ and $b$ were  respectively $(2,3)$ and $(3,4)$ for the 2 bands. These transformations allow us to make multi-band images that are not trivially linearly related and yet largely share the same morphological properties and center.

 Then, each individual band of the truth data set is convolved with a random Moffat PSF ($0.6 \leq FWHM \leq 1$ arcsecond, $2 \leq \beta \leq 5$, $-0.9 \leq g1,g2 \leq 0.9$, $\sigma_{g1,g2} = 0.1$, $\mu_{g1,g2} = 0$)  and random Gaussian noise was added using \texttt{addNoiseSNR} function in \texttt{GalSim} ($0.01\leq \sigma \leq 0.1$).  The pixel scale of the PSF of band 1 is adopted to be the pixel scale for the other two band PSFs and when rendering the truth images. For the entire data set this varies between $0.088$ and $0.255$ pixels/arcseconds. This makes our data unrealistically heterogeneous, but should not affect the conclusions.

For each image we calculate the total signal-to-noise ratio (SNR) defined as 
\begin{equation}
    SNR = \sqrt{\frac{{\sum_{\rm pixels}(\rm truth \circledast PSF})^{2}}{\sigma^2}}.
\end{equation}
We find that the overall SNR in our set spans 0.8-350 with a mode at around SNR of 60. 

The entire dataset is divided into training and test sets where the training set contains $90,000$ images, while the test set contains $10,000$ images. Each image is sized $35\times35\times3$ and normalized to $[0,1]$ by dividing by the overall maximum pixel value of the noisy and truth images so that the colors are preserved. This normalization is necessary due to large natural dynamic range in the astronomical images, where fluxes can vary by orders of magnitude, significantly more than in the case of ordinary photographic images. This has several important ramifications. The first is that the noisy objects contribute approximately the same to the loss function as bright objects. We have not attempted to correct for this factor in the loss function. Another obvious downside is that the noise level and even the offset is not uniform from image to image. Given that we know the noise level rms in the renormalized image, but never pass this information to the network, we know that our approach must be suboptimal at some level. Nevertheless, the network seems to adapt fine to the varying noise levels.  After the network outputs its predictions, these predictions are rescaled back by the same factor which allows a direct comparison with the ground truth to evaluate the performance of the network accurately.

\section{Results}

\subsection {Implementation and comparison models}
In our framework, each band-wise deconvolution module contains $m=3$ residual blocks.  During the first phase of the training process, the model is trained for $200$ epochs with a learning rate of $10^{-4}$ and a batch size of $128$. In the second phase, we fine-tune the model for $30$ epochs and the learning rate is set to $10^{-5}$. In both phases, Adam optimizer \citep{kingma2014adam} is adopted to update the parameters in the model. The weights for moment-related loss are manually as following: $\lambda_{0, 1, 2} = 0.01$, $\lambda_3 = 0.002$ and $\lambda_{4, 5} = 0.001$. Those parameters are chosen to 1) keep the dominance of the $\ell_1$-loss and 2) scale the loss terms to be comparable.

In order to evaluate the performance of our model we train 4 different models:
\begin{itemize}
    \item The fiducial model, where we train as described in the text;
    \item The \emph{$\ell_1$ loss only} model, where training is not fine-tuned with custom loss function;
    \item The \emph{Average PSF} model, where the training and testing datasets remain the same, but instead of fitting actual per-image PSF to the network, we freeze the PSF to its average shape. This allows us to explore the information gain coming from knowledge of the actual PSF
    \item The \emph{Gaussian regularization} model where Wiener deconvolution formula is replaced with a different regularization as described in Section \ref{sec:shaperec}.
\end{itemize}

\subsection {Evaluation}

We start by noting that the output of a neural network is a nominally deconvolved and denoised image. As opposed to a classical Wiener filter, which can suppress the noise variance, but cannot distinguish between the noise and the data, the DWDN builds an internal model of how galaxies look like and the result will therefore be a nominally noise-less image generated from this internal representation that is consistent with the data. This image is of course not truly noiseless in the sense that there will be a difference between the reconstructed image and the ground truth, but it is noiseless in the sense that the pixel values outside the extent of the object are pinned to zero. We have explicitly checked that those values are at the relative level of about $10^{-3}$. We can therefore treat neural network image output as noiseless in terms of calculating the quantities of interest.

In most tests we split the sample into three categories by the value of $M_{00}$ as a proxy for SNR, and refer to the resulting sub samples as \emph{Low}, \emph{Medium} and \emph{High} SNR.  Although the mapping is not perfect, these bins correspond to SNR ratios that fall approximately into 0.8-70, 70-200 and 200-350 regions respectively. 

\begin{figure}[ht]
    \centering
    \subfloat[Sample result with low SNR]{\includegraphics[width=0.9\textwidth]{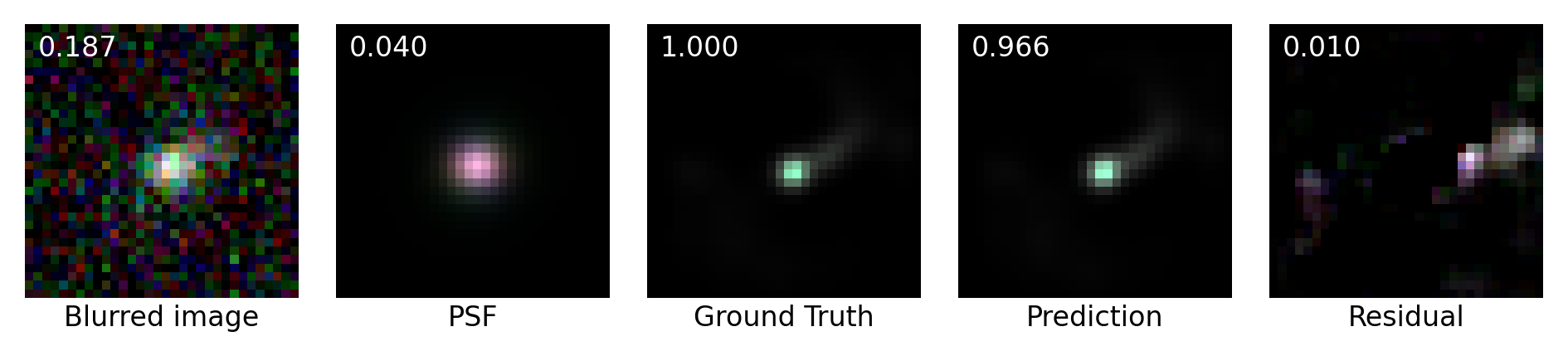}} \\
    \subfloat[Sample result with medium SNR]{\includegraphics[width=0.9\linewidth]{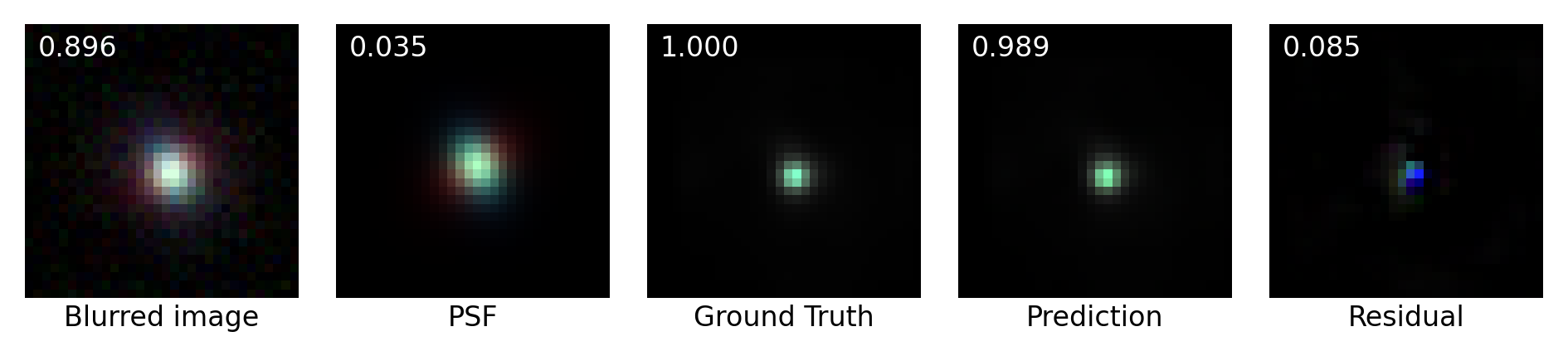}} \\
    \subfloat[Sample result with high SNR]{\includegraphics[width=0.9\linewidth]{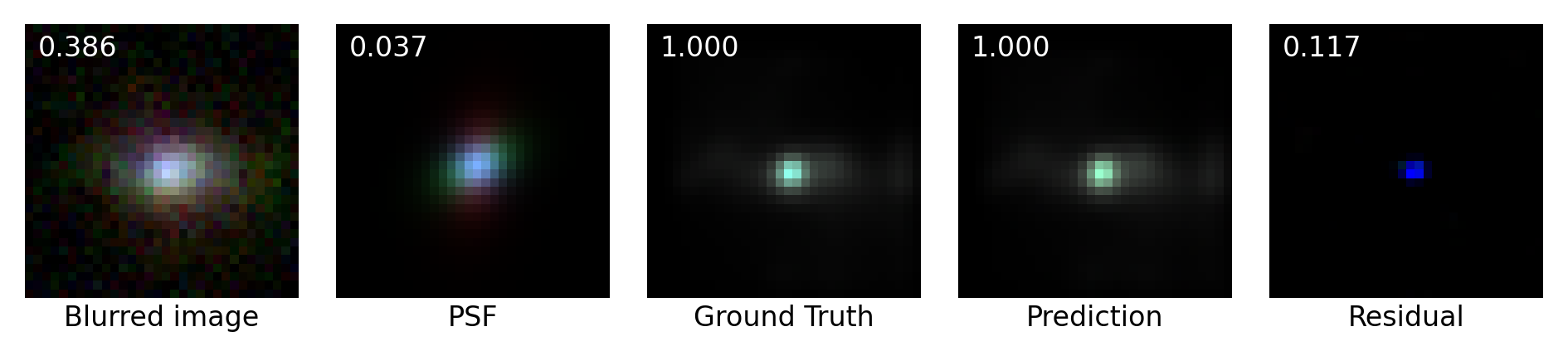}} \\
    \subfloat[Sample result of a galaxy with morphological structure]{\includegraphics[width=0.9\linewidth]{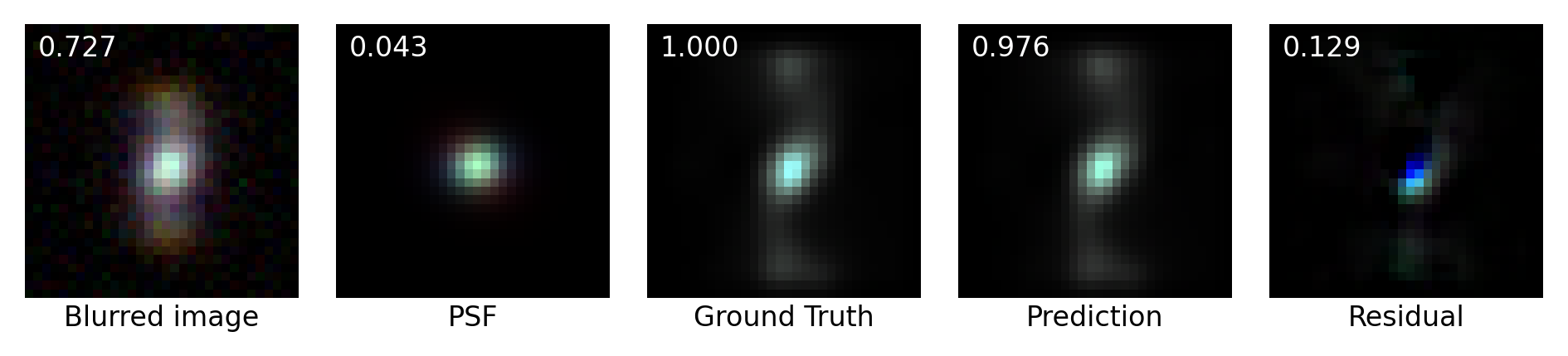}} \\
    \subfloat[Sample result of a galaxy with rich morphological structure]{\includegraphics[width=0.9\linewidth]{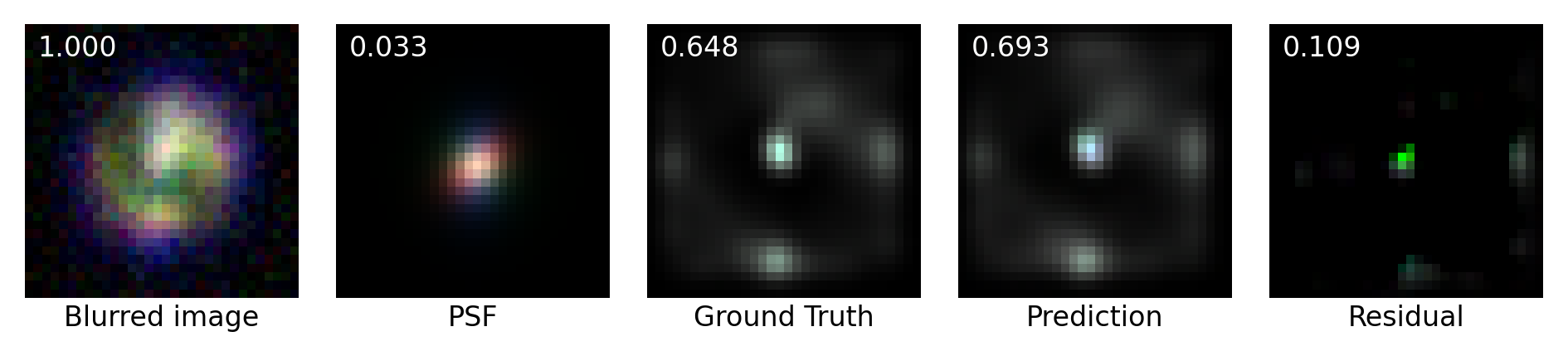}} \\
    \caption{Figure showing the visualization of typical results. The columns correspond to i) noisy input image; ii) PSF assumed to be perfectly known; iii) ground truth that we are trying to recover; iv) Neural network's guess of the ground truth and v) the difference between ground truth and the prediction. Numbers in the upper left corner denotes the relative normalization of the flux scale. }
\label{fig:visualization}
\end{figure}

Fig. \ref{fig:visualization} visualizes the performance of the trained model on the testing set. We select one example each from  with low, medium and high SNR sub samples and show as cherry picked examples in which galaxies have richer morphological structure.  From the results, our deblurring model can recover the morphological and color information for images under different conditions, even for images with low fluxes. It can also be observed that the model performs better on images with higher fluxes. We also find that the residual image is devoid of noise outside the core of the object.  We will analyse these results in quantitative detail in the following sections.

We first use some standard metrics such as the peak signal-to-noise ratio (PSNR) and the structural similarity index (SSIM) to evaluate the quality of images recovered from our model compared with the true images. 
PSNR 
is the ratio between the maximum of the true image and the mean squared error of the true and recovered images in logarithmic decibel scale. The expression for PSNR is in Eq. \ref{eq:PSNR}. 
SSIM~\citep{wang2004image} evaluates the similarity between two images based on quantities related to visible structures in the image such as mean ($\mu_x$, $\mu_y$), variance ($\sigma_x$, $\sigma_y$) and covariance ($\sigma_{xy}$). The SSIM is formulated as Eq.~\ref{eq:SSIM} in which $c_1$ and $c_2$ are small constants included to avoid instability. We use the default settings $c_1=(k_1L)^2$ and $c_2=(k_1L)^2$ where $k_1=0.01$ and $k_2=0.03$. and $L$ is the data range. A higher PSNR or SSIM value represents better image recovery.

\begin{equation}
    {\rm PSNR(dB)} = 20\cdot \log_{10}({\rm MAX})-10\cdot \log_{10}({\rm MSE}) \label{eq:PSNR}
\end{equation} 

\begin{equation}
    {\rm SSIM} = \frac{(2\mu_x\mu_y+c_1)(2\sigma_{xy}+c_2)}{(\mu_x^2+\mu_y^2+c_1)(\sigma_x^2+\sigma_y^2+c_2)} \label{eq:SSIM}
\end{equation}

The evaluation of our fiducial model and the three comparison models are listed in Table \ref{tab:psnr_ssim}, including the mean and median values for the testing set. The best results are from the model with accurate PSF and moment loss. Specifically, by comparing  the first and the third columns in the table, we can conclude that the moment-related loss terms help improve the performance, albeit modestly,  as reflected by the higher PSNR and SSIM values. The results from the models with accurate PSF and average PSF also demonstrate that the quality of the prediction will get degraded if an accurate PSF is not provided for each image again with  modest improvements.

\begin{table*}
\centering
\begin{tabular}{c|cc|cc|cc|cc}
  \hline \hline
  & \multicolumn{2}{c|}{Accurate PSF} & \multicolumn{2}{c|}{Accurate PSF ($\ell_1$ loss)}   & \multicolumn{2}{c|}{Average PSF} &\multicolumn{2}{c}{Gaussian regularization}\\
  \hline
  & Mean  & Median  & Mean  & Median & Mean  & Median & Mean  & Median \\
\hline 
PSNR & 43.2157 & 42.7412  & 43.6333 & 42.8883 & 42.5585  & 41.8337
    & 43.0713 & 42.4740 \\
1-SSIM  & 0.0222 & 0.0095 & 0.0211 & 0.0090  & 0.0253 & 0.0124  & 0.0229 & 0.0105\\
  \hline \hline
\end{tabular}
\caption{\label{tab:psnr_ssim} PSNR(dB) and SSIM }
\end{table*}

\begin{table*}[htbp]
\centering
\begin{tabular}{c|cc|cc|cc|cc}
  \hline \hline
  & \multicolumn{2}{c|}{Accurate PSF } & \multicolumn{2}{c|}{Accurate PSF ($\ell_1$ loss)}   & \multicolumn{2}{c|}{Average PSF} & \multicolumn{2}{c}{Gaussian regularization} \\
  \hline
  & Mean  & RMS & Mean & RMS & Mean & RMS & Mean & RMS \\
\hline 
$M_{00}$ 
& 1.8963  & 86.6546
& 3.0573 & 86.8562 
& 3.3452  & 99.5131 
& 2.6432 & 89.1245  
\\
$\left<x\right>$ 

& -0.0053 & 0.1629
& 0.0009 & 0.1403 
& -0.0102 & 0.1687 
& -0.0052 & 0.1752 
\\
$\left<y\right>$ 

& -0.0011 & 0.1618
& 0.0007 & 0.1363 
& -0.0142 & 0.1698 
& 0.0026 & 0.1648  
\\

$e_{1}$ 

& 0.0011 & 0.0575
& -0.0003 & 0.0574 

& -0.0056 & 0.0847 
& 0.0003 & 0.1214  
\\
$e_{2}$ 

& 0.0016 & 0.0681
& -0.0012 & 0.0635 

& 0.0033 & 0.0964 
& -8.9499e-5 & 0.0708 
\\
$\|\boldsymbol{e}\|$ 

& -0.0092 & 0.0612
& -0.0020 & 0.0568 

& -0.0180 & 0.0851 
& -0.0084 & 0.1924 
\\
$s$ 

& 0.3697 & 3.4984
& -0.1884 & 3.3433 
& 0.3164 & 4.0902 
& 0.1824 & 3.6217  
\\
  \hline \hline
\end{tabular}
\caption{\label{tab:moment_difference} Difference between the quantities of interest for the predicted images and the ground truth. }
\end{table*}

\begin{table*}[htbp]
\centering
\begin{tabular}{c|cc|cc|cc}
  \hline \hline
  & \multicolumn{2}{c|}{Low SNR} & \multicolumn{2}{c|}{Medium SNR} & \multicolumn{2}{c}{High SNR} \\
  \hline
  & Mean  & RMS & Mean & RMS & Mean & RMS\\
\hline 
$M_{00}$ 
& 5.1935  & 31.1791 
& 5.3326 & 60.3967
& -1.3547 & 134.2164 
\\
$M_{00}/M_{00\_true}$ 
& 0.4722  & 1.8063 
& 0.0210 & 0.2415
& 0.0021 & 0.1345 
\\
$\left<x\right>$ 
& 6.2498e-5 & 0.1453   
& -0.0018 & 0.1410
&  0.0044 & 0.1428
\\
$\left<y\right>$ 
& 0.0024 & 0.1470
& 0.0016 & 0.1409
& -0.0018 & 0.1196
\\
$e_{1}$ 
& -0.0005  & 0.0714
& -0.0009  & 0.0553
& 0.0004 & 0.0417
\\
$e_{2}$ 
& -0.0019 & 0.0766
& -0.0012 & 0.0617
&  -0.0005 & 0.0494
\\
$\|\boldsymbol{e}\|$ 
& -0.0006 & 0.0692 
& -0.0047 & 0.0543
& -0.0006 & 0.0442
\\
$s$ 
& -0.1994 & 2.9974
& -0.1113 & 3.7088
& -0.2544 & 3.2855
\\
$s/s_{true}$ 
& 0.0042 & 0.2112
& 0.0038 & 0.1403
& -0.0020 & 0.0778
\\
  \hline \hline
\end{tabular}
\caption{\label{tab:moment_difference_high} Difference between the  quantities of interest for the predicted images and the ground truth split by SNR for the model trained by ($\ell_1$ loss).}
\end{table*}

\begin{table*}[htbp]
\centering
\begin{tabular}{c|cc|cc|cc|cc}
  \hline \hline
  & \multicolumn{2}{c|}{Accurate PSF } & \multicolumn{2}{c|}{Accurate PSF ($\ell_1$ loss)}   & \multicolumn{2}{c|}{Average PSF} & \multicolumn{2}{c}{Gaussian regularization}\\
  \hline
  & Mean & RMS & Mean & RMS & Mean & RMS & Mean & RMS \\
\hline 
band 3 - band 1 
& 0.0056 & 0.0346 
& 0.0030 & 0.0350 

& 0.0032 & 0.0341 
& 0.0020 & 0.0316 
\\
band 1 - band 2 
&-0.0021 & 0.0654 
&-0.0028 & 0.0610 

& 0.0018 & 0.0676 
& 0.0032 & 0.0631 
\\
band 2 - band 3  
& -0.0035 & 0.0704 
& -0.0002 & 0.0682 

& -0.0049 & 0.0751 
& -0.0052 & 0.0704 
\\
  \hline \hline
\end{tabular}
\caption{\label{tab:color}Astronomical magnitude difference between two bands.}  
\end{table*}

\subsubsection{Recovery of astronomical quantities}

\begin{figure*}[ht]
    \centering
    \subfloat[$\log(M_{00})$]{\includegraphics[width=0.33\textwidth]{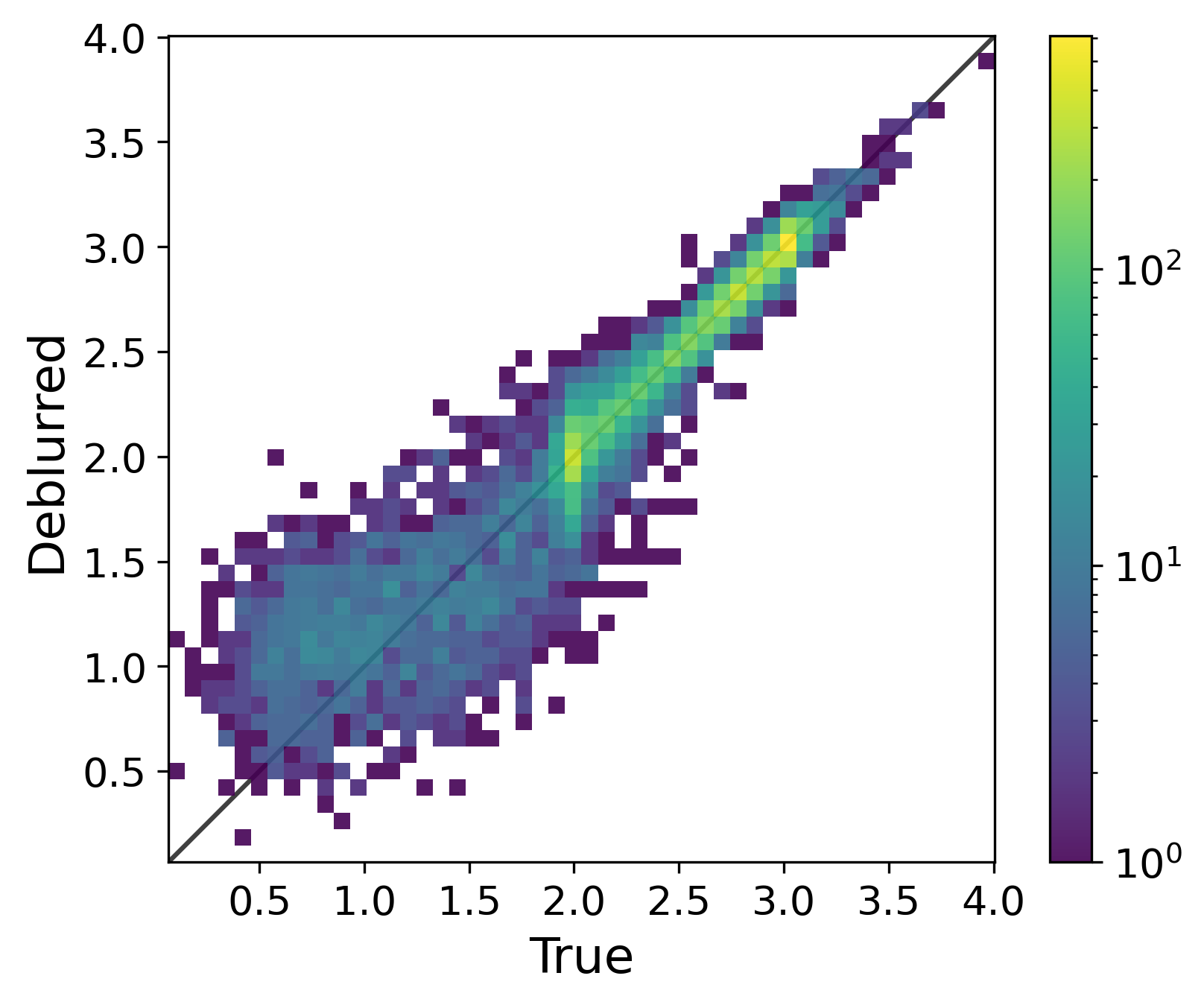}}
    \subfloat[$\bar{x}$]{\includegraphics[width=0.33\linewidth]{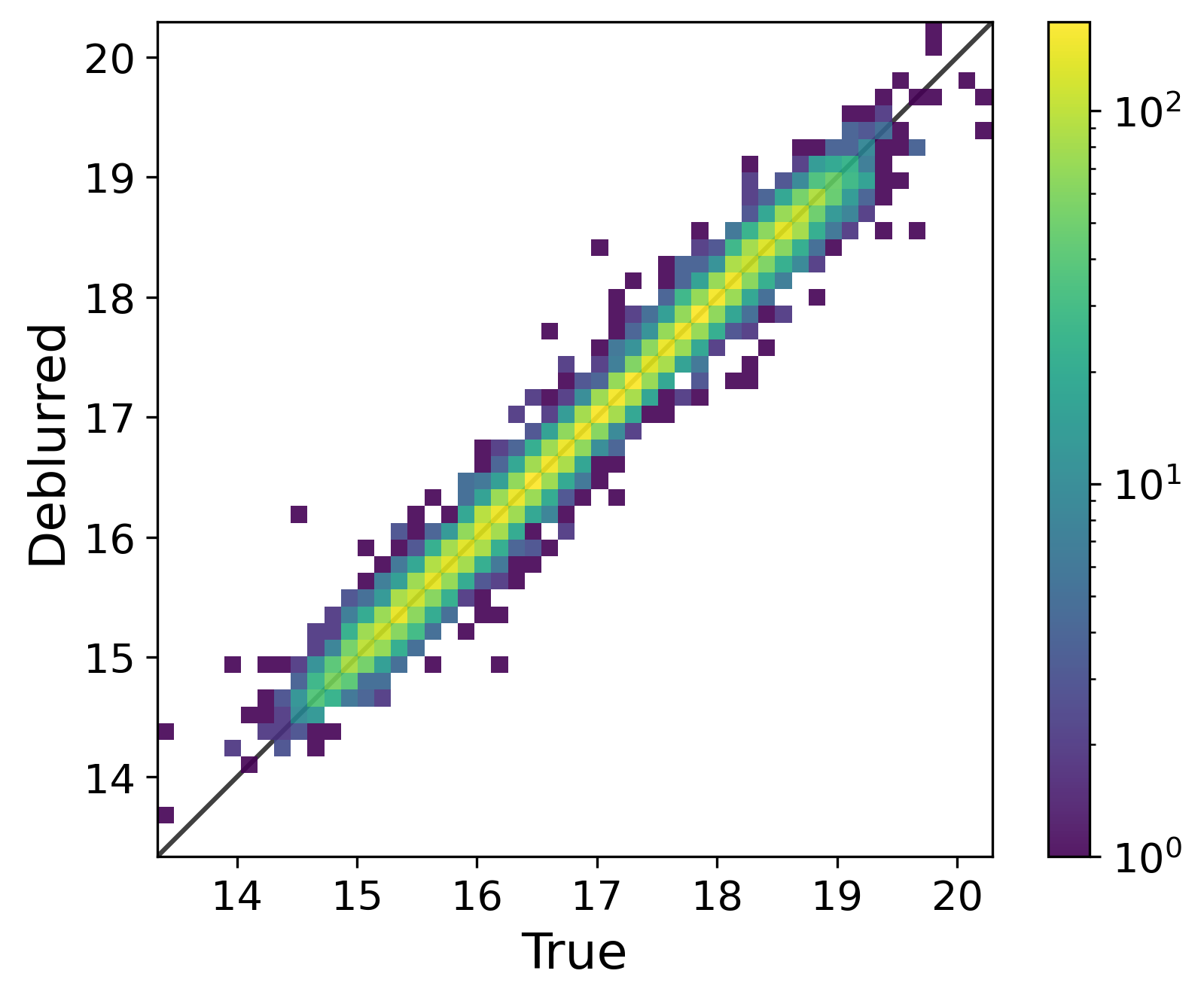}} 
    \subfloat[$\bar{y}$]{\includegraphics[width=0.33\linewidth]{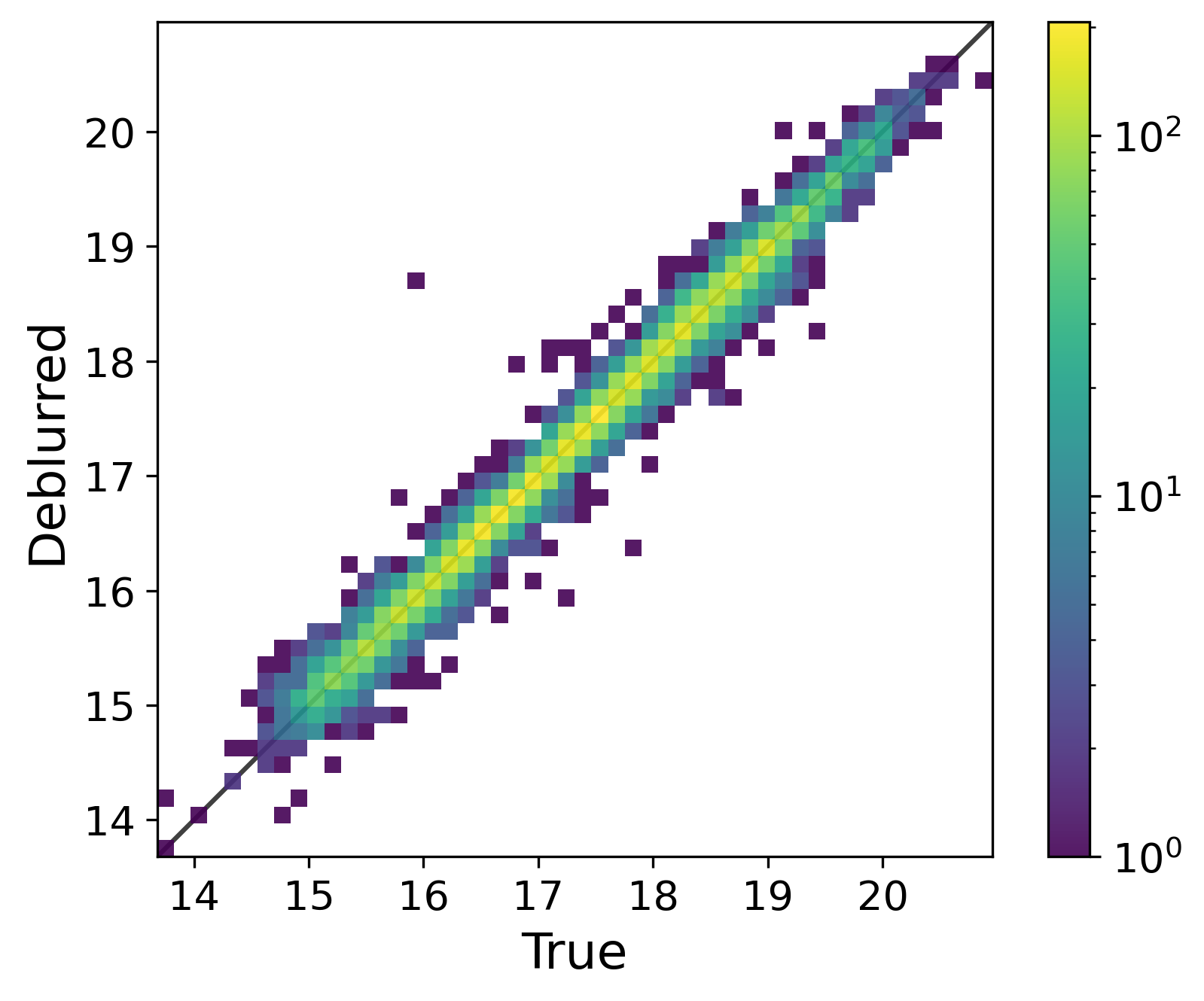}} \\
    \subfloat[$e_{1}$]{\includegraphics[width=0.33\linewidth]{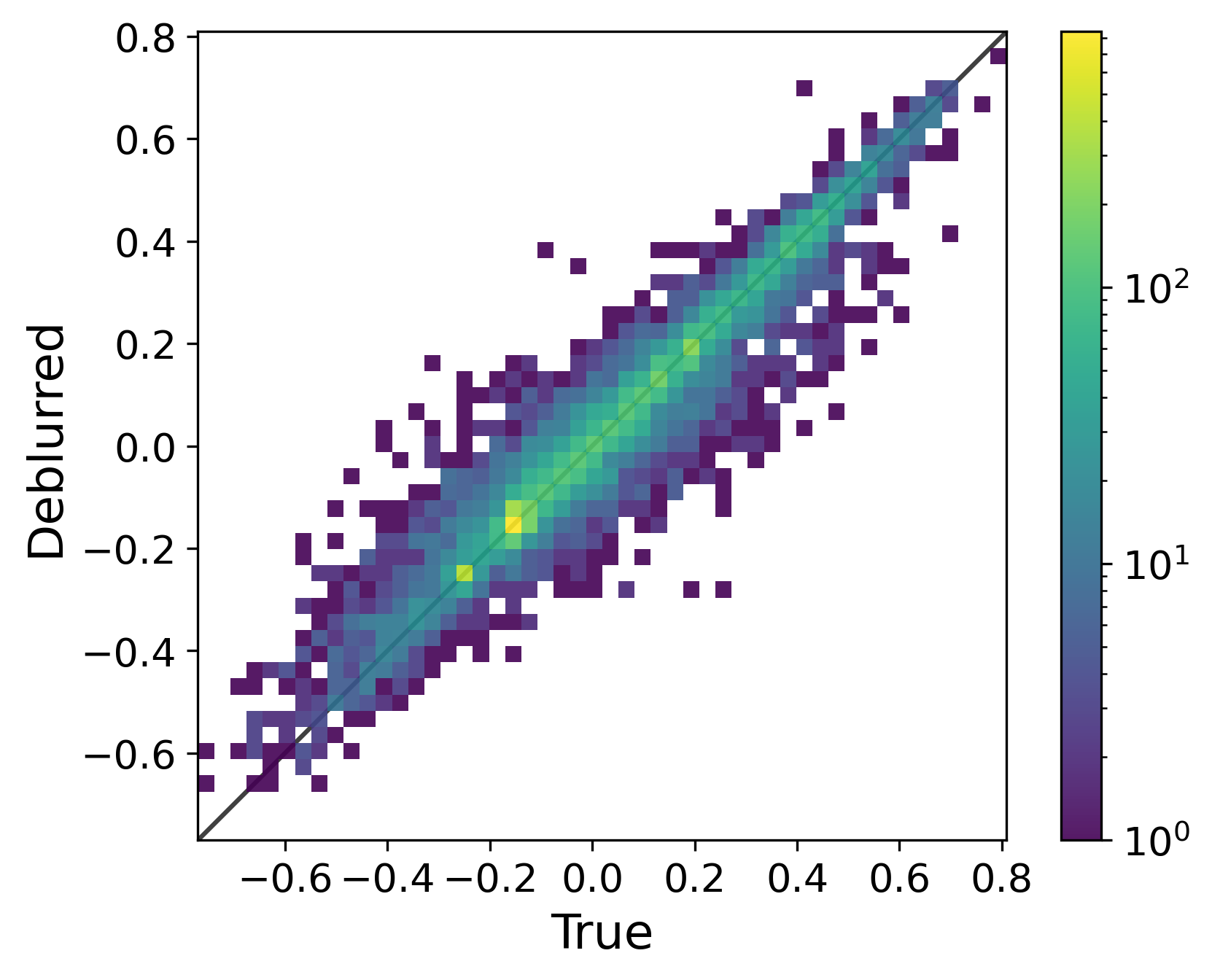}}
    \subfloat[$e_{2}$]{\includegraphics[width=0.33\linewidth]{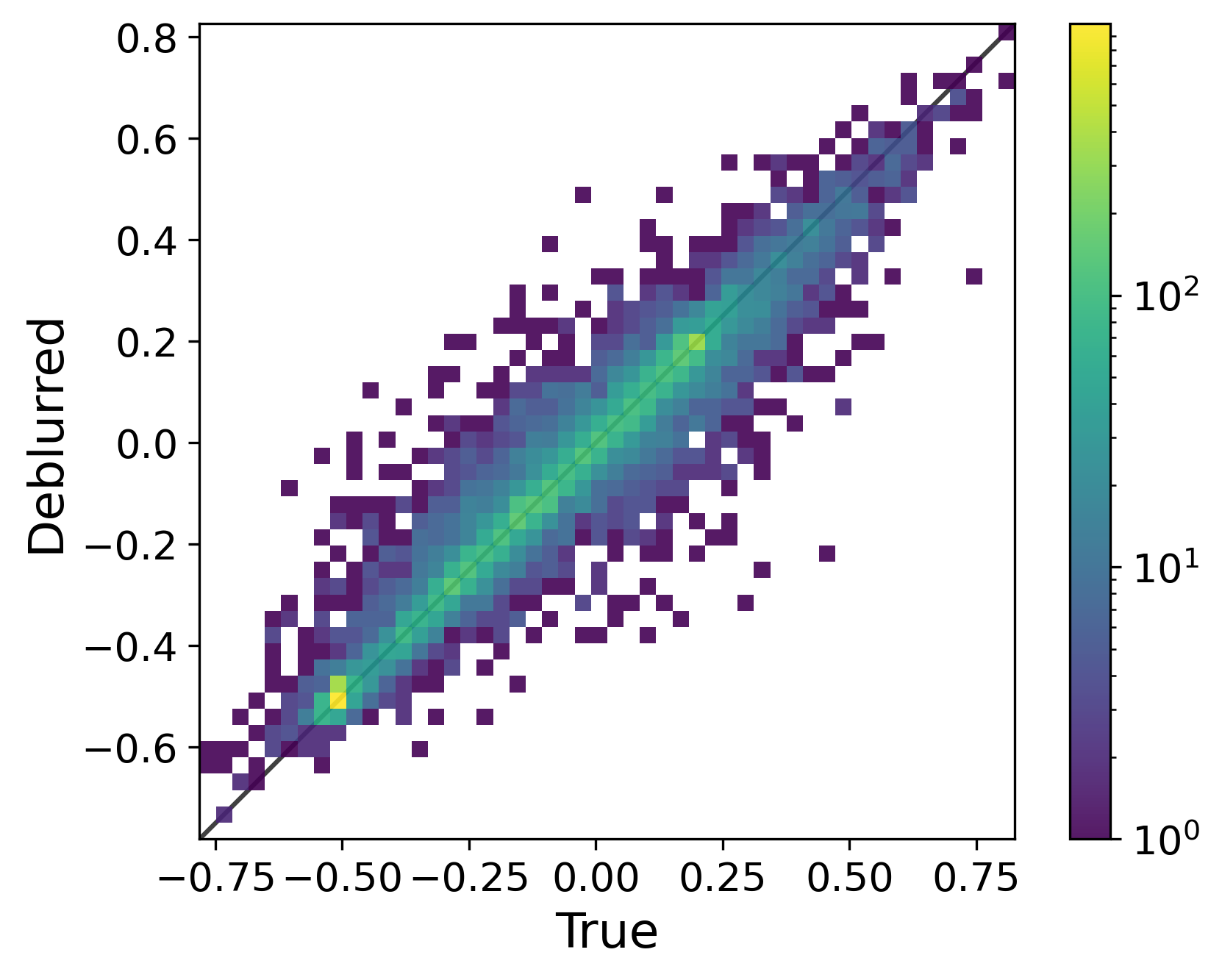}}
    \subfloat[$s$]{\includegraphics[width=0.33\linewidth]{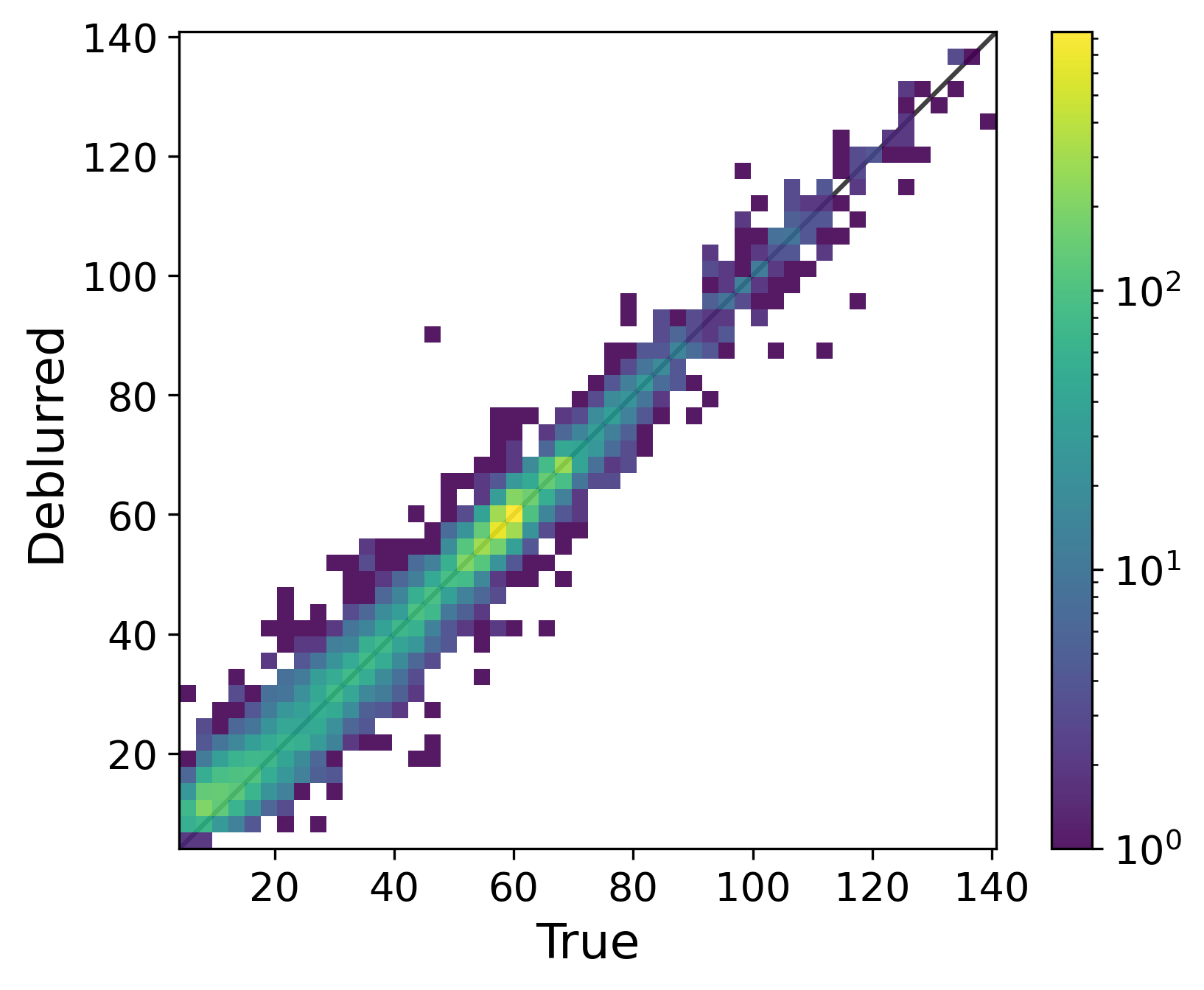}}
\caption{2-d histograms showing quantity recovery from the model trained with accurate PSF, the quantity names are indicated in the labels. The $x$ axis was calculated on the true image and the $y$ axis on the recovered image. The density of points is represented with color. Note that the color scale is logarithmic to bring out the outliers. For the total intensity we also plot the histogram in the log of the value due to its large dynamic range. }
\label{fig:moment_orig}
\end{figure*}

\begin{figure*}
    \centering
    \subfloat[$e_{1}$ for low SNR]{\includegraphics[width=0.3\textwidth]{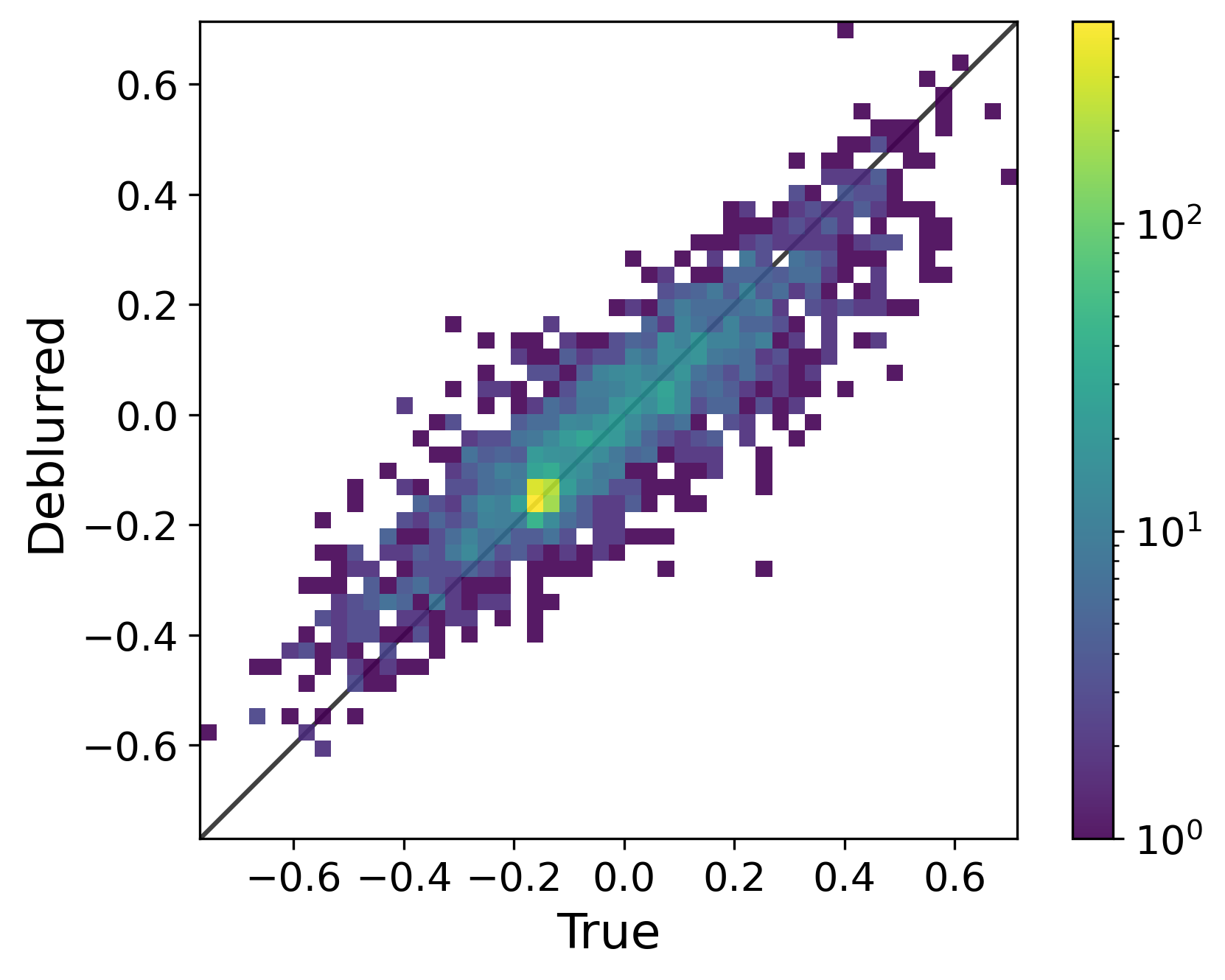}}
    \subfloat[$e_{1}$ for medium SNR]{\includegraphics[width=0.3\linewidth]{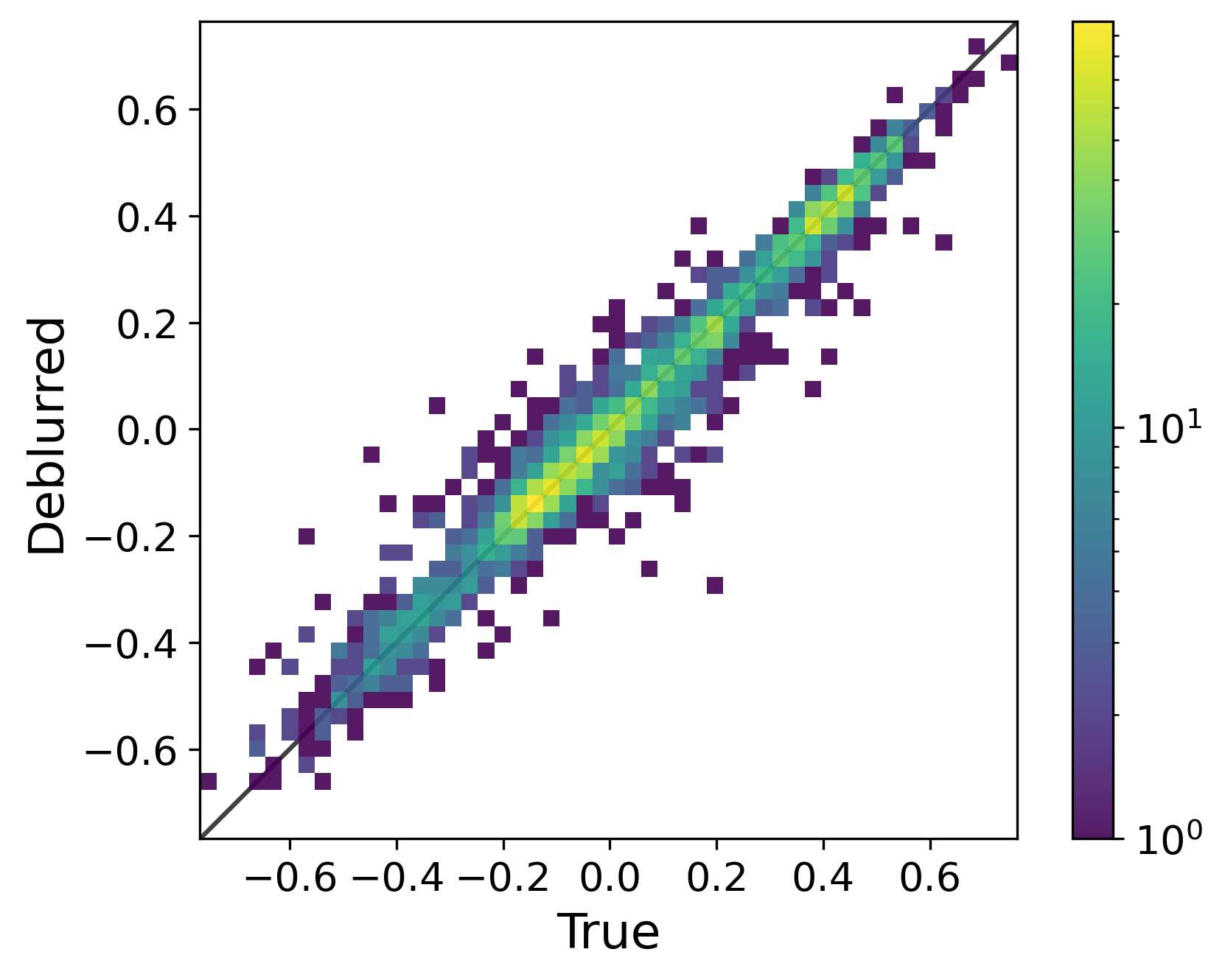}} 
    \subfloat[$e_{1}$ for high SNR]{\includegraphics[width=0.3\linewidth]{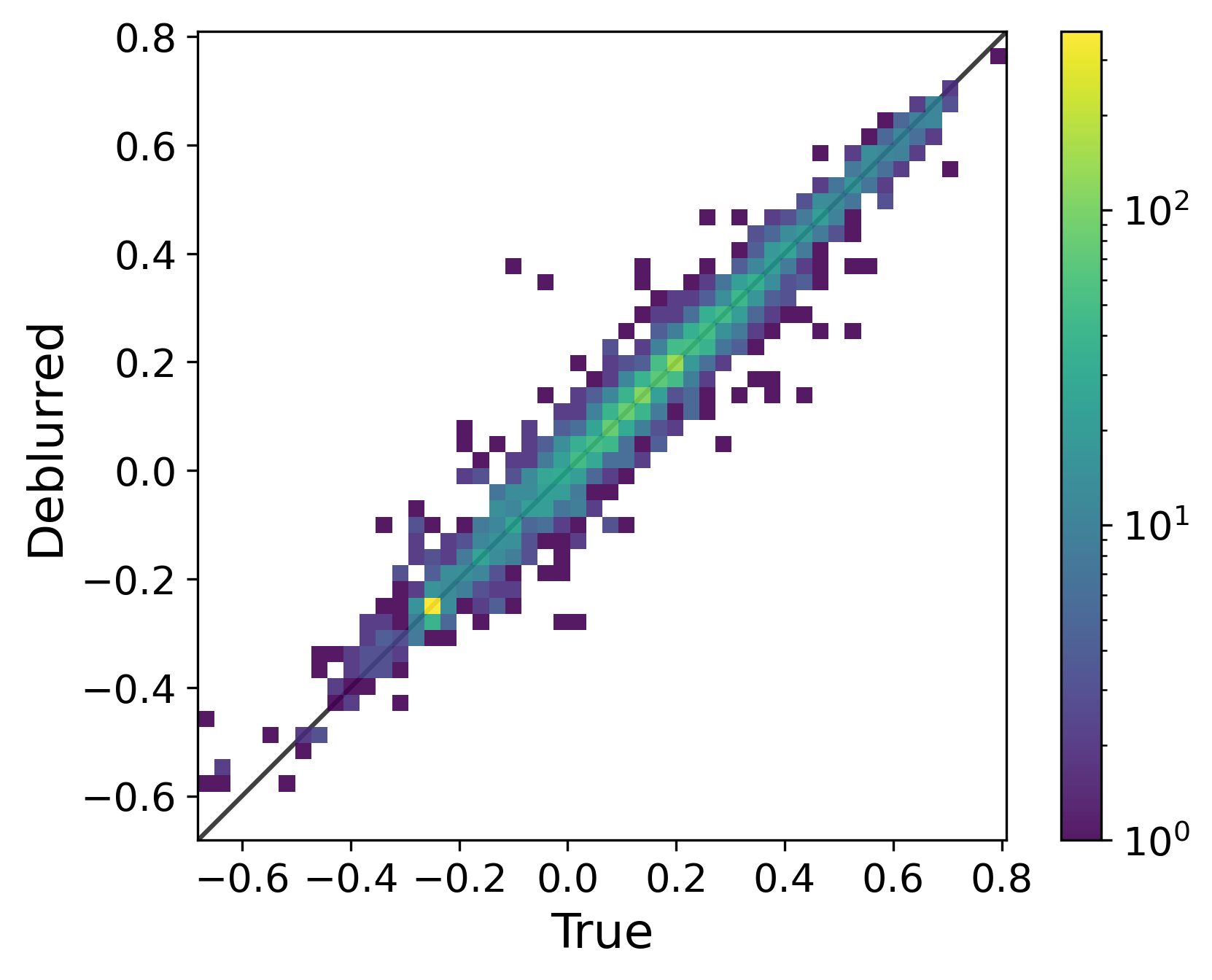}} \\
    \subfloat[$e_{2}$ for low SNR]{\includegraphics[width=0.3\linewidth]{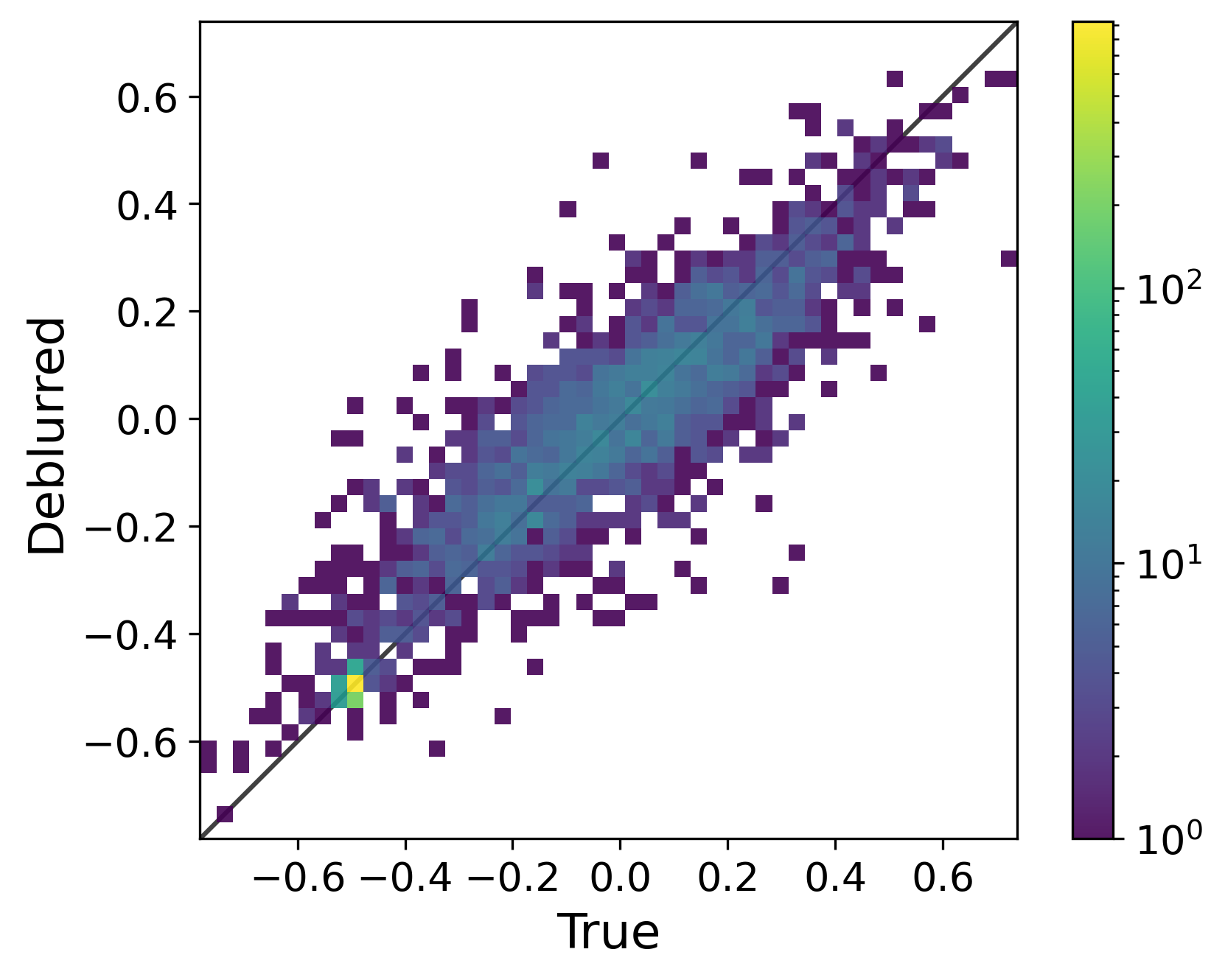}} 
    \subfloat[$e_{2}$ for medium SNR]{\includegraphics[width=0.3\linewidth]{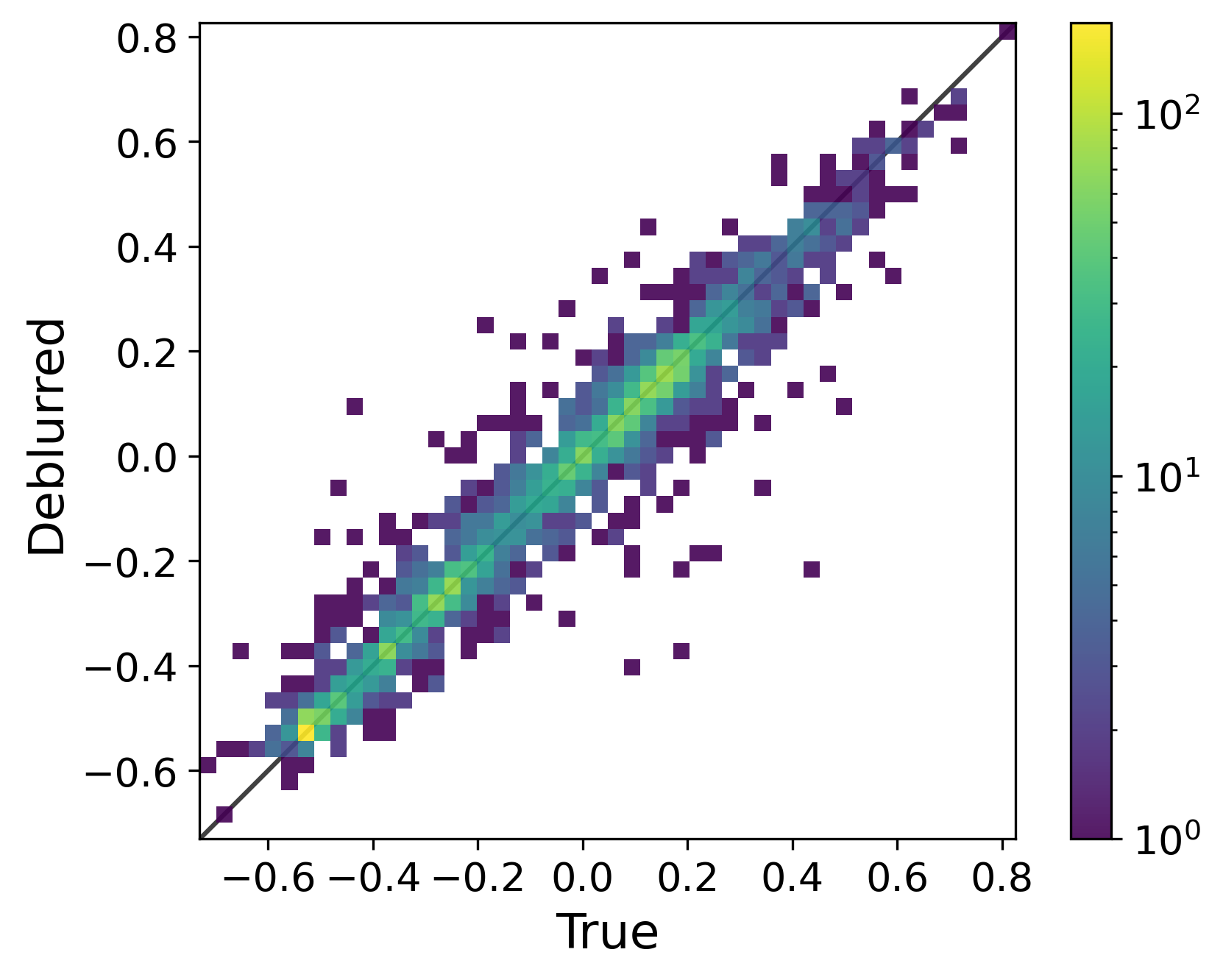}}
    \subfloat[$e_{2}$ for high SNR]{\includegraphics[width=0.3\linewidth]{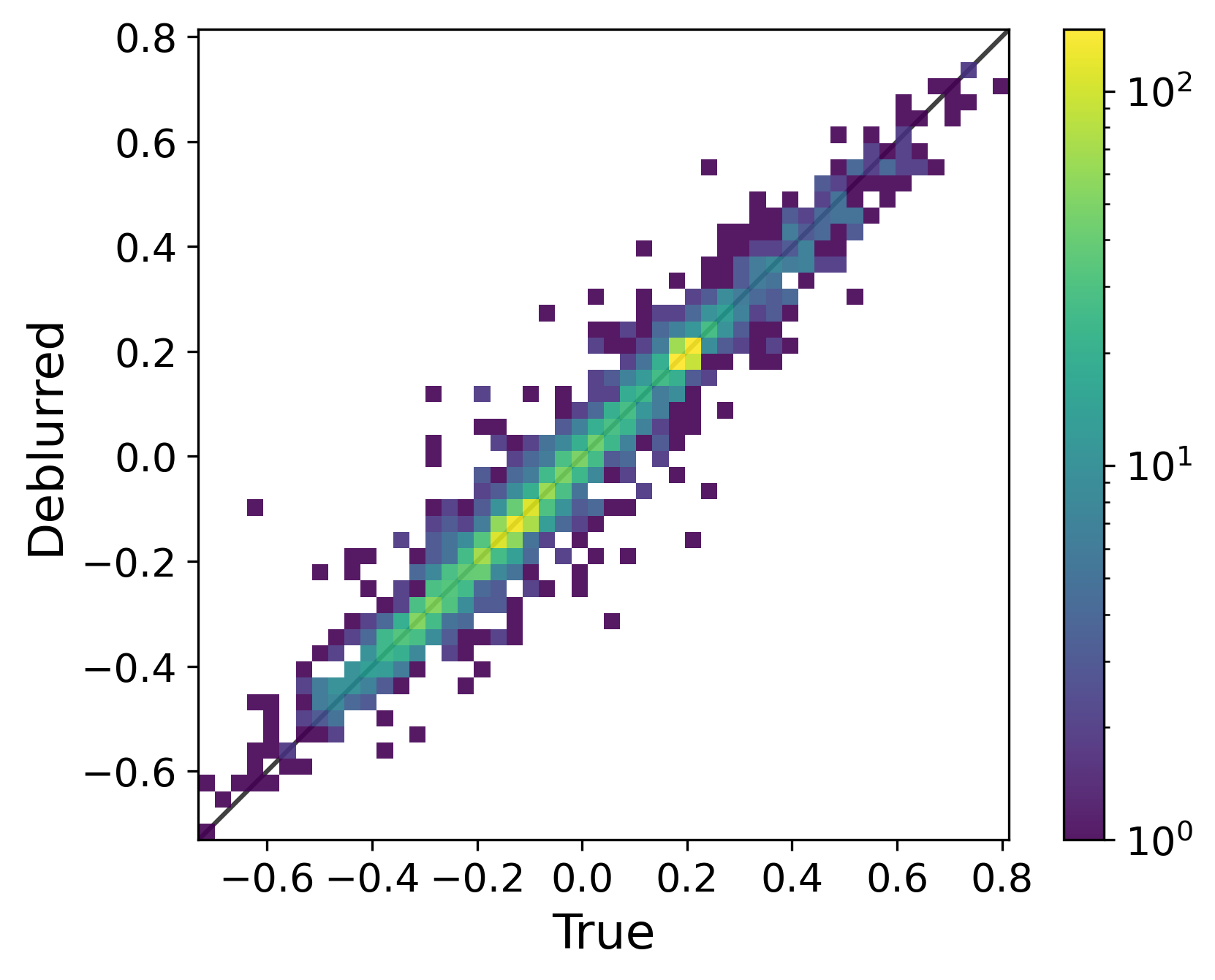}} \\
    \caption{2D histograms showing ellipticity recovery split by SNR for the fiducial model.}
\label{fig:lmh_orig}
\end{figure*}

Table \ref{tab:moment_difference} presents the difference between the prediction from the model and the ground truth image for the astronomical quantities of interest. These values are averaged over the three available bands. For our fiducial network, the relevant scatter plots can also be found in Figure \ref{fig:moment_orig}. We find that across the noise levels, the networks correctly recover the quantities of interest. The results nicely scatter around the $x=y$ line with no obvious bias for the total flux, first moments (astrometric positions) and second moments. Ellipticity suffers from a small bias, where the network tends to make objects rounder. To understand this effect better, we have split objects by SNR and re-plotted the ellipticity histograms in Figure \ref{fig:lmh_orig}. Here we see that high SNR objects have their shapes recovered without bias, while there is an emergence of ``multiplicative'' bias for low SNR where measured ellipticity components are systematically low. 

We see that the our fiducial network performs worse than the the network with a tuned loss function, despite loss functions being optimized for precisely the quantities of interest.Our investigations provide two possible explanations. The first possibility is that the adaptive moments are measuring a different quantity. In Figure \ref{fig:visualization} we see that some objects are significantly extended and in this case the adaptive moments would focus on the brightest object and suppress other structures. In fact, remaking Table \ref{tab:moment_difference} calculated using adaptive moments instead of naive moments does show numerical differences, but qualitatively the story is largely unchanged. The other possibility is that we are over-training. Indeed, we have explicitly confirmed that while moment contribution to the loss function continues to fall during training, the results on the test data do not improve. We discuss this further in the conclusions.  

\begin{figure}
    \centering
    \subfloat[$e_{1}$ for low SNR]{\includegraphics[width=0.3\textwidth]{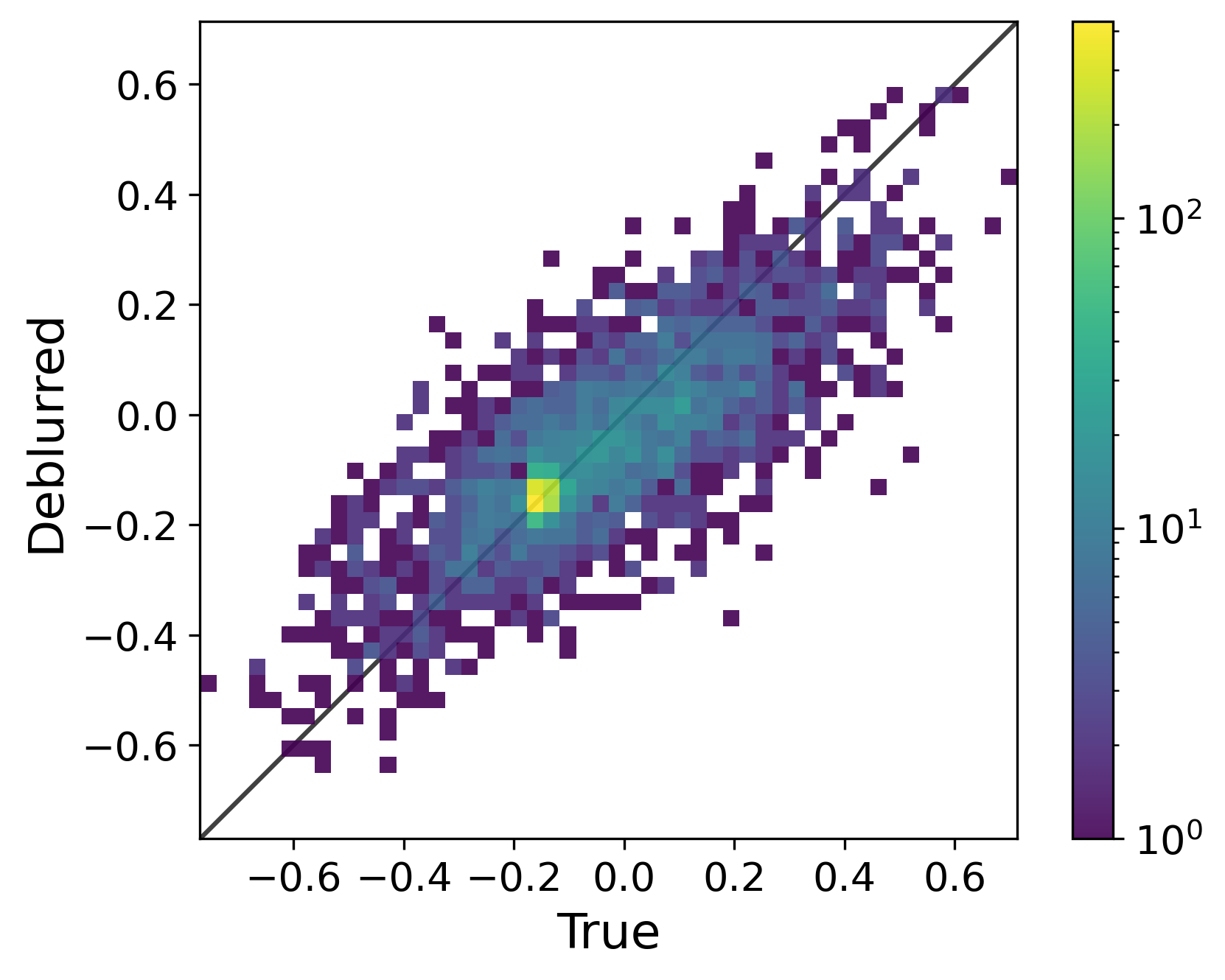}}
    \subfloat[$e_{1}$ for medium SNR]{\includegraphics[width=0.3\linewidth]{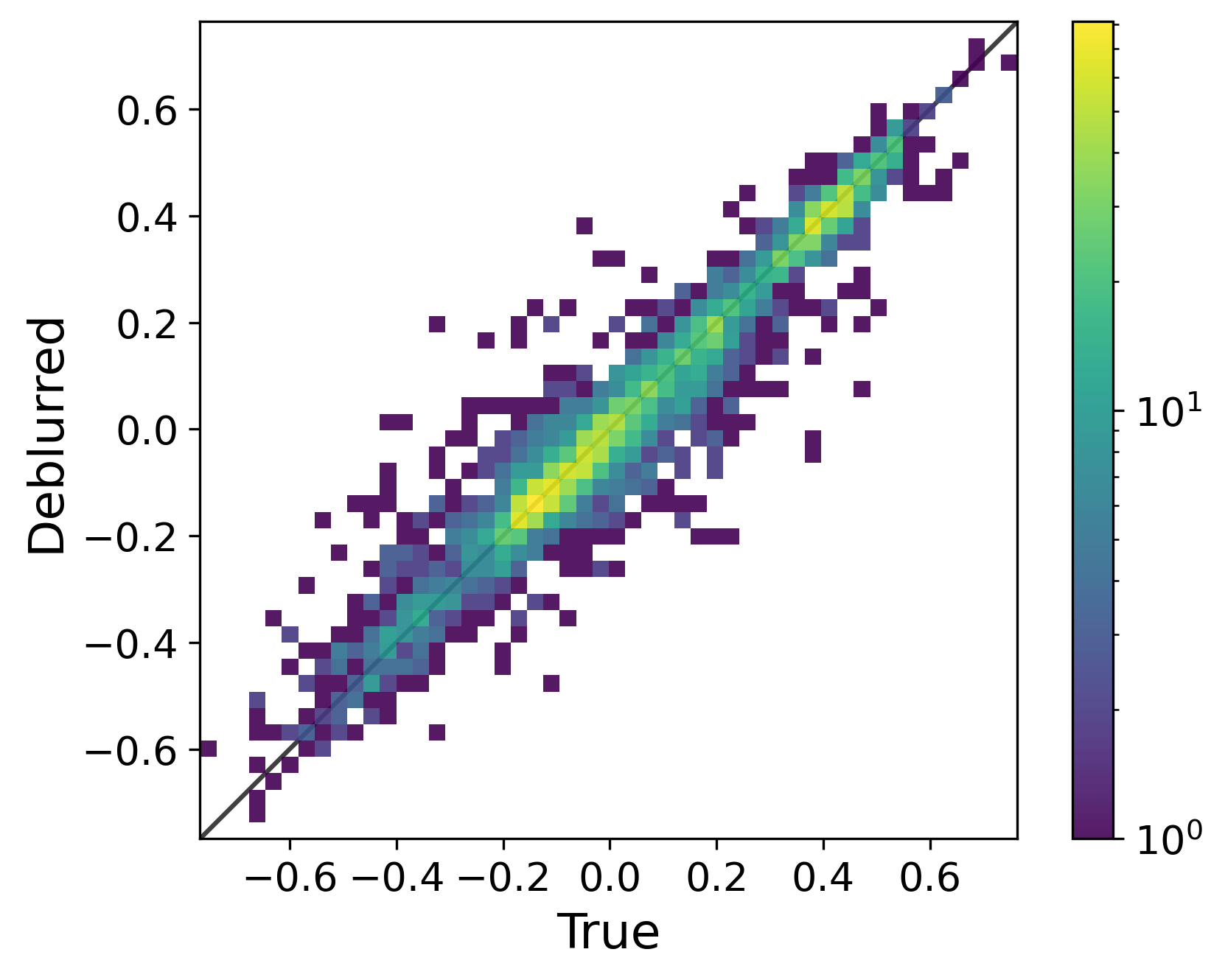}} 
    \subfloat[$e_{1}$ for high SNR]{\includegraphics[width=0.3\linewidth]{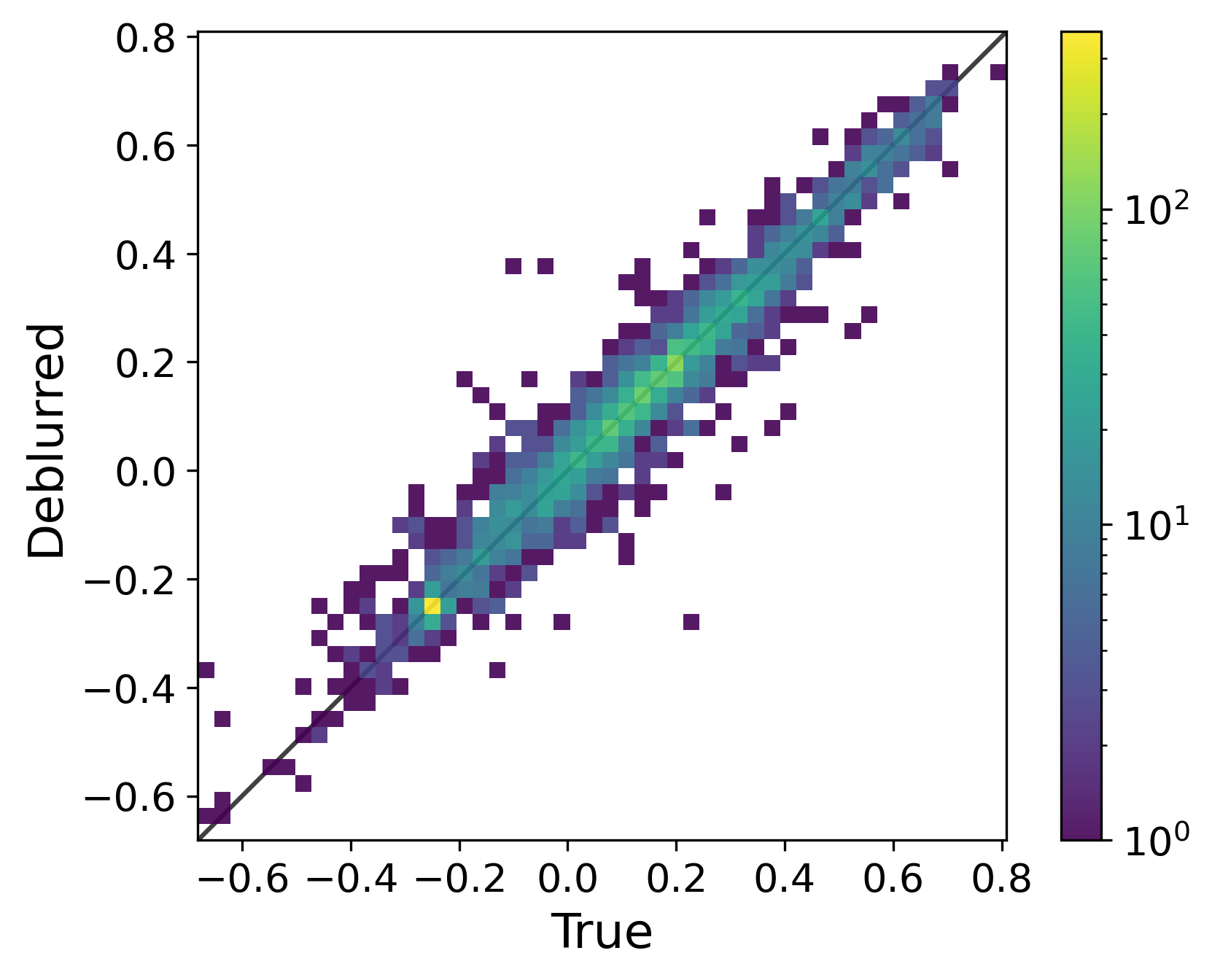}} \\
    \subfloat[$e_{2}$ for low SNR]{\includegraphics[width=0.3\linewidth]{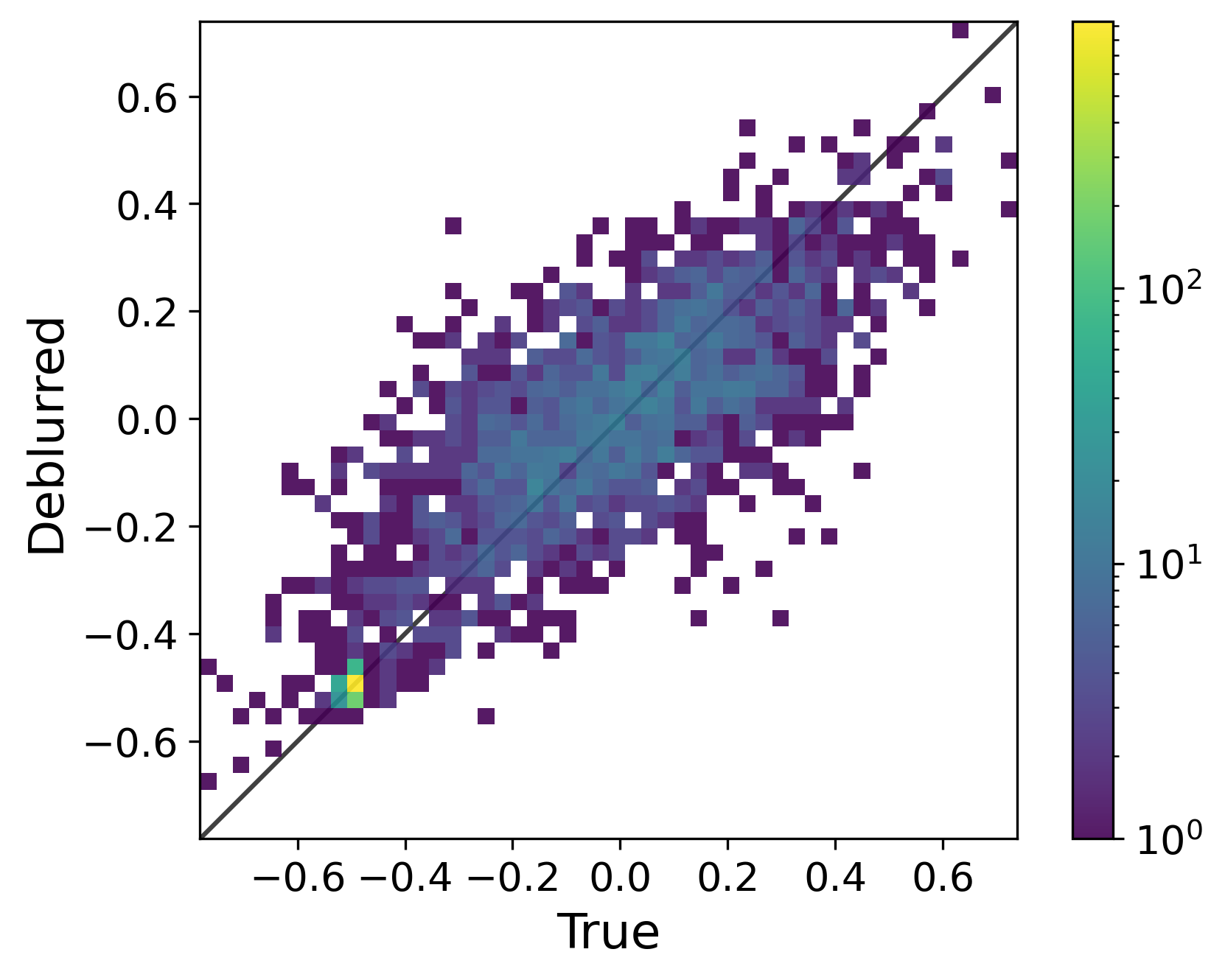}} 
    \subfloat[$e_{2}$ for medium SNR]{\includegraphics[width=0.3\linewidth]{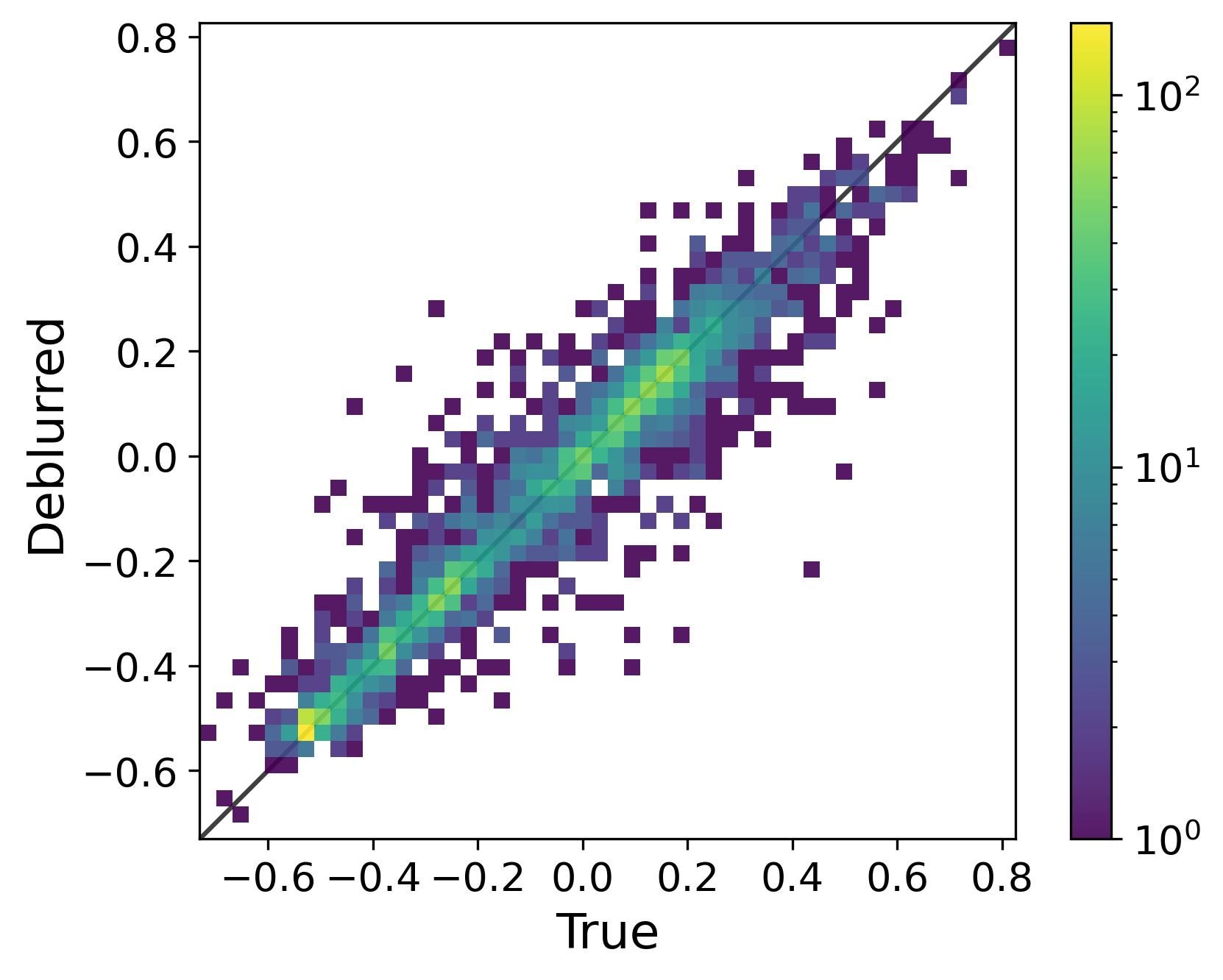}}
    \subfloat[$e_{2}$ for high SNR]{\includegraphics[width=0.3\linewidth]{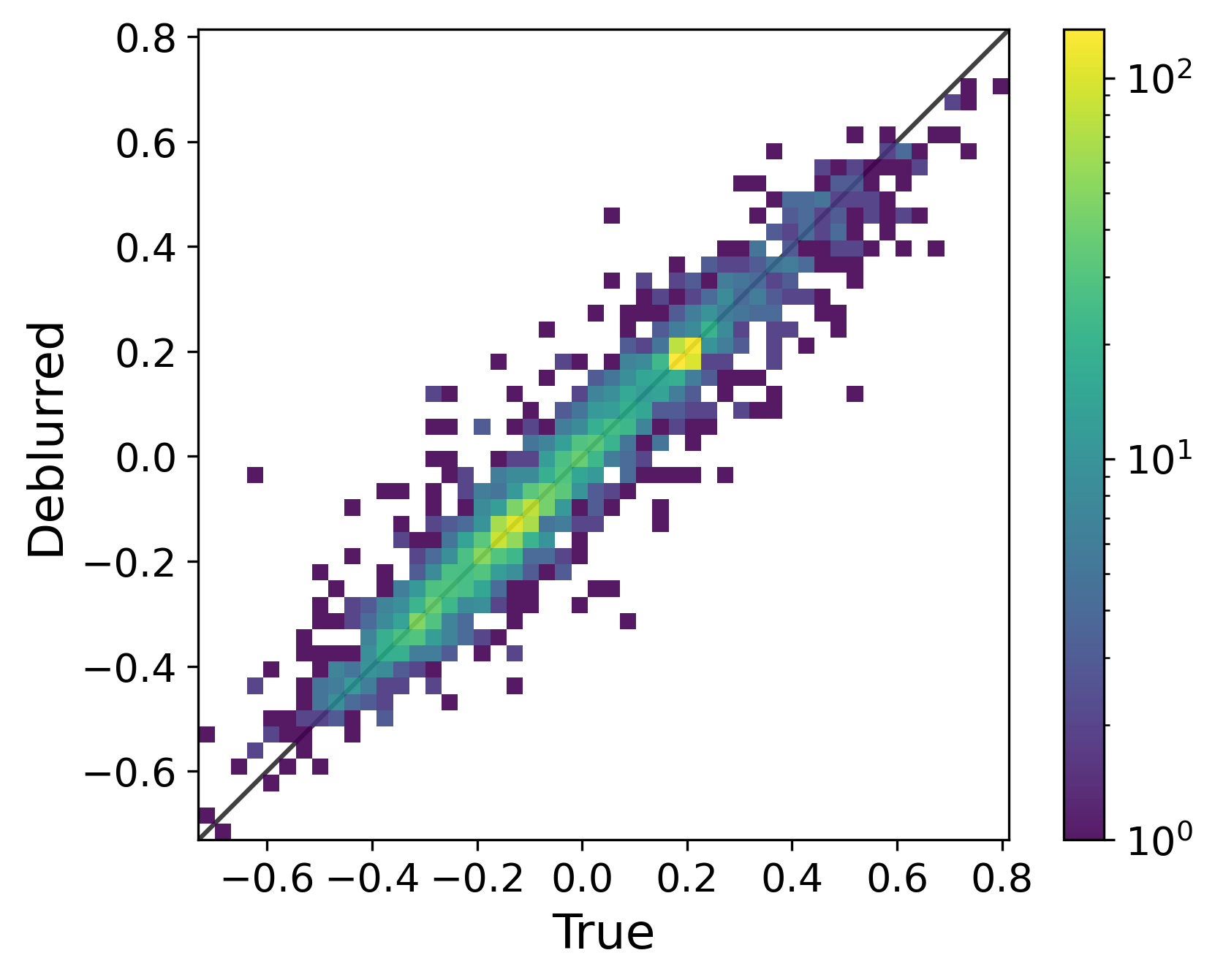}} \\
    \caption{2D histograms showing ellipticity recovery split by SNR for the model trained with average PSF.}
\label{fig:lmh_average}
\end{figure}

Next we want to study how much information we are getting by using a per-band known PSF. The equivalent of Figure \ref{fig:lmh_orig} for the model with average PSF is plotted in Figure \ref{fig:lmh_average}. We see essentially the same structure, but the scatter around the mean is somewhat higher. This is confirmed in Table \ref{tab:moment_difference}, where RMS (root mean square) errors when using average PSF (2nd column) are systematically larger by about $5-10\%$ compared to those of the fiducial network for both ellipticity and second moments. We also find that the error on the astrometrics are not affected by the knowledge of the true PSF -- this makes sense because our PSFs are symmetric with zero mean and therefore using a wrong PSF to deconvolve the image is not going to affect its central position.

To investigate this further, we turn to color recovery. We define object colors to be the logarithm of the ratio for $M_{00}$ between two channels written as
\begin{equation}
band \ 3 - band \ 1 = -2.5\times\log{\frac{(M_{00})_{band \ 3}}{(M_{00})_{band \ 1}}}.
\end{equation}
The color thus defined corresponds approximately to the color schema used by astronomers, which is the magnitude difference between two bands.  Results are shown in Table \ref{tab:color}. We see $10\%$ improvement in color recovery when the accurate rather than the average PSF is used. The reason for this is a band dependent PSF, which can artificially introduce the color structure into the output object. This is also illustrated in Figure \ref{fig:visualization}, where we see that the object in the blurred image is often of a different color than the object in the ground truth image.

\begin{figure}
    \centering
    \includegraphics[width=0.49\linewidth]{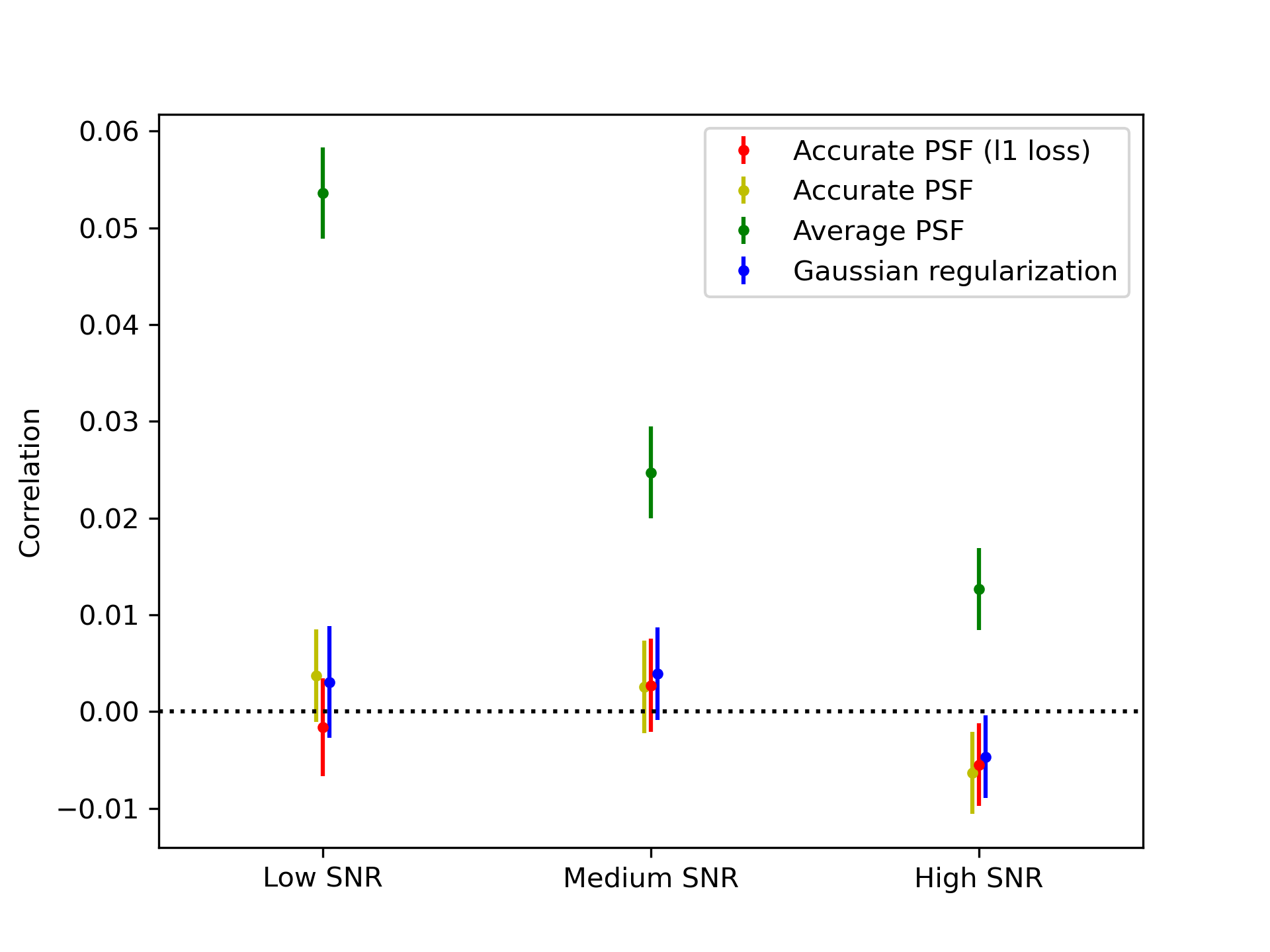}
    \includegraphics[width=0.49\linewidth]{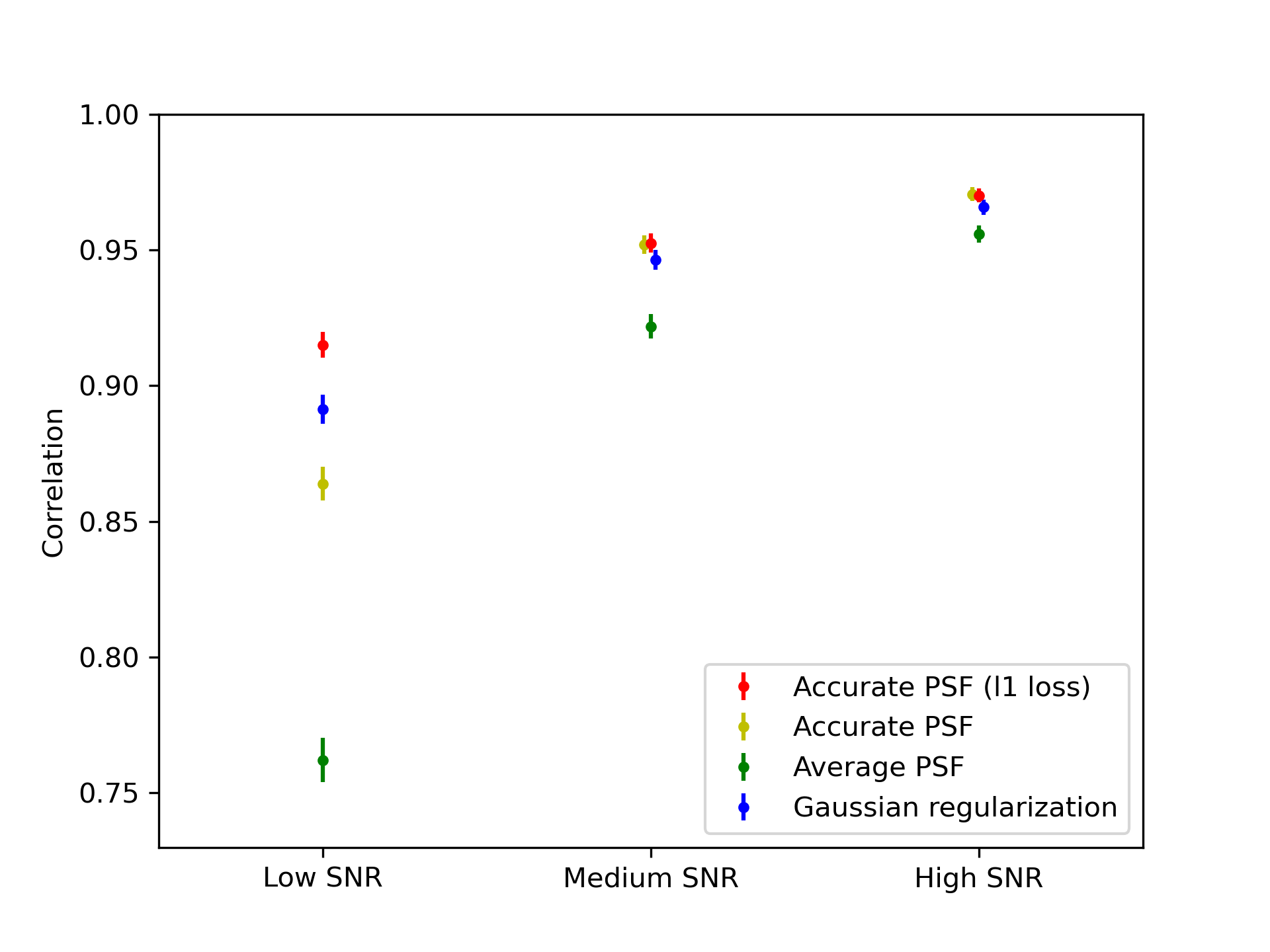}
    \caption{Plots showing the cross correlations of the network output -- the left panel shows the cross correlation of  the ellipticities with the PSF and the right panel shows that with the true image ellipticities.}
    \label{fig:shape_corr}
\end{figure}

\subsubsection{Recovery of sizes}

In Tables \ref{tab:moment_difference} and \ref{tab:moment_difference_high} we also show the results for the galaxy size $s$, (split  by the model in the former and by SNR in the latter). The size $s$ as defined here  (see Equation \ref{eq:moments}) is measured in pixels and since larger galaxies generally have higher SNR, it also correlates with the SNR. We see that the error on $s$ is largely independent of SNR, but what fractional error there is decreases with increasing SNR, as expected.  Note that the sign in the mean between $s$ and $s/s_{\rm true}$ can be different since the implicit weighting is different. The same result is plotted in the bottom right planel of \ref{fig:moment_orig}.

In standard linear Wiener deconvolution, one expects that the correlation length of the recovered image increases as we push towards fainter end, since the wave number at which the image becomes noise dominated and the filter starts to suppress the noise increases. It is therefore natural that the linear Wiener filter exhibits a fainter-fatter effect. We see no such effect here, since the bias in size is always considerably smaller than the variance and doesn't exhibit  a clear trend with the SNR. Since the Wiener filter is internal to the DWDN, our interpretation is that the refinement steps correct for any over-smoothing present in the Wiener filter. 

\subsubsection{Recovery of shapes and its biases}
\label{sec:shaperec}
Next we study the performance of our network in removing the subtle biases that remain in the predictions. In particular, we want to study the remaining contamination by the PSF shape. Our main motivation is to understand whether such methods could ever be used for weak lensing applications, where residual contamination with the PSF leads to the so-called additive bias (see \citealt{RM2018} and references therein for an in-depth review). In short, if the shape of the galaxy remains correlated with the PSF after deconvolution, then this PSF shape will leak into shear estimation and lead to shape correlations that can mimic real weak gravitational signal. 

The true galaxy ellipticities are uncorrelated with the PSF shapes, but since the observed image is a convolution of the PSF with the true shape, the resulting predicted image ellipticities might still correlate with the PSF shape. On the other hand, if the measurement is perfect, the predicted ellipticities should correlate perfectly with the true ones. To quantify this effect, we define the following correlations
\newcommand{\xitrue}{\xi_{\rm true}}
\newcommand{\xipsf}{\xi_{\rm PSF}}

\begin{eqnarray}
\xipsf &=&  \frac{\left<\boldsymbol{e}_{\rm PSF}\cdot \boldsymbol{e}_{\rm pred}\right>}{\left<|\boldsymbol{e}_{\rm PSF}| |\boldsymbol{e}_{\rm pred}|\right>},\\
\xitrue &=& \frac{\left <\boldsymbol{e}_{\rm true}\cdot \boldsymbol{e}_{\rm pred}\right>}{\left<|\boldsymbol{e}_{\rm true}| |\boldsymbol{e}_{\rm pred}|\right>},
\end{eqnarray}
where averages are calculated by simply averaging the data in the testing set. Note that since individual ellipticities of either the PSF, the true galaxy image or the predicted galaxy image can reach zero, we average the numerator and denominator separately. In an ideal case, we expect the predicted image to remove any trace of the PSF from the resulting image, so $\xipsf=0$. However, this could also be achieved by making all objects round (i.e. $\boldsymbol{e}_{\rm pred}=0$ for every object). Therefore we also look at the correlation with the true value, which should be, in the ideal case perfect, i.e  $\xitrue=1$. 

Results of this experiment are plotted in Figure \ref{fig:shape_corr}. We see that in the case of average PSF, there are significant levels of residual correlation with the input PSF that gets more pronounced towards lower SNR.  Note that the task given to the average PSF model is without a solution : it is mathematically impossible to separate the effect of intrinsic shear from that of the PSF if all we have available is the resulting image. Therefore it effectively tries to make a best guess assuming that the PSF is circular (which it is, in average).  We see that our fiducial case (accurate PSF) performs  correctly at all SNR levels. Most importantly, there is no evidence of correlation with the PSF and the correlation with true shear is significantly higher that in the case of average PSF. 

Again we find that the simple $\ell_1$ loss works best across noise levels and especially in the low SNR regime.

In another experiment that we performed with smaller galaxies with exponential profiles and Gaussian PSFs,  we did find residual sensitivity to the PSF. While our accurate PSF model performed better, the result of the cross-correlation with PSF was inconsistent with zero.  For completeness, we present our solution to that problem that can be viewed as bias-hardening of the network with minimal impact on the results.  The idea is to explore another version of deconvolution in which we replace the Wiener filter Equation \ref{eq:wiener1} with a modified term, which we refer to as ``Gaussian Regularization'':
\begin{equation}
    \WD^{\rm GaussReg} = \frac{P_{s} (k)}{P_{s}(k)+P_{n}(k) G^{-2}(k)} \PSF^{-1}(k),
    \label{eq:wiener2}
\end{equation}
In other words, the PSF is linearly deconvolved from the noisy image, but the resulting image is then regularized with a circular Gaussian. In the original formulation of the Wiener filter, modes are anisotropically regularized, taking into account the fact that the anisotropic PSF has destroyed more information in some directions than the others. However, the result is that the PSF shape can sneak back into the resulting image. The Gaussian regularization tries to prevent this at the expense of formally decreased optimality, i.e. the resulting process is now manifesting a worse than minimum variance solution. We chose a Gaussian with a  size close to the size of the PSF and have explicitly demonstrated that the results are largely insensitive to the precise choice of this kernel, if the network is re trained.

This method is plotted in blue in Figure \ref{fig:shape_corr}. We see that the results are very similar to the accurate PSF results, but for other datasets this method might provide a better protection against PSF leakage.

In our main results presented in Tables \ref{tab:moment_difference} and \ref{tab:moment_difference_high}, we see that the Gaussian regularization performs essentially as well as the fiducial model and therefore it should be applied whenever independence from the PSF takes priority over raw image fidelity.

\begin{figure}[ht]
    \centering
    \subfloat[low SNR]{\includegraphics[width=0.3\textwidth]{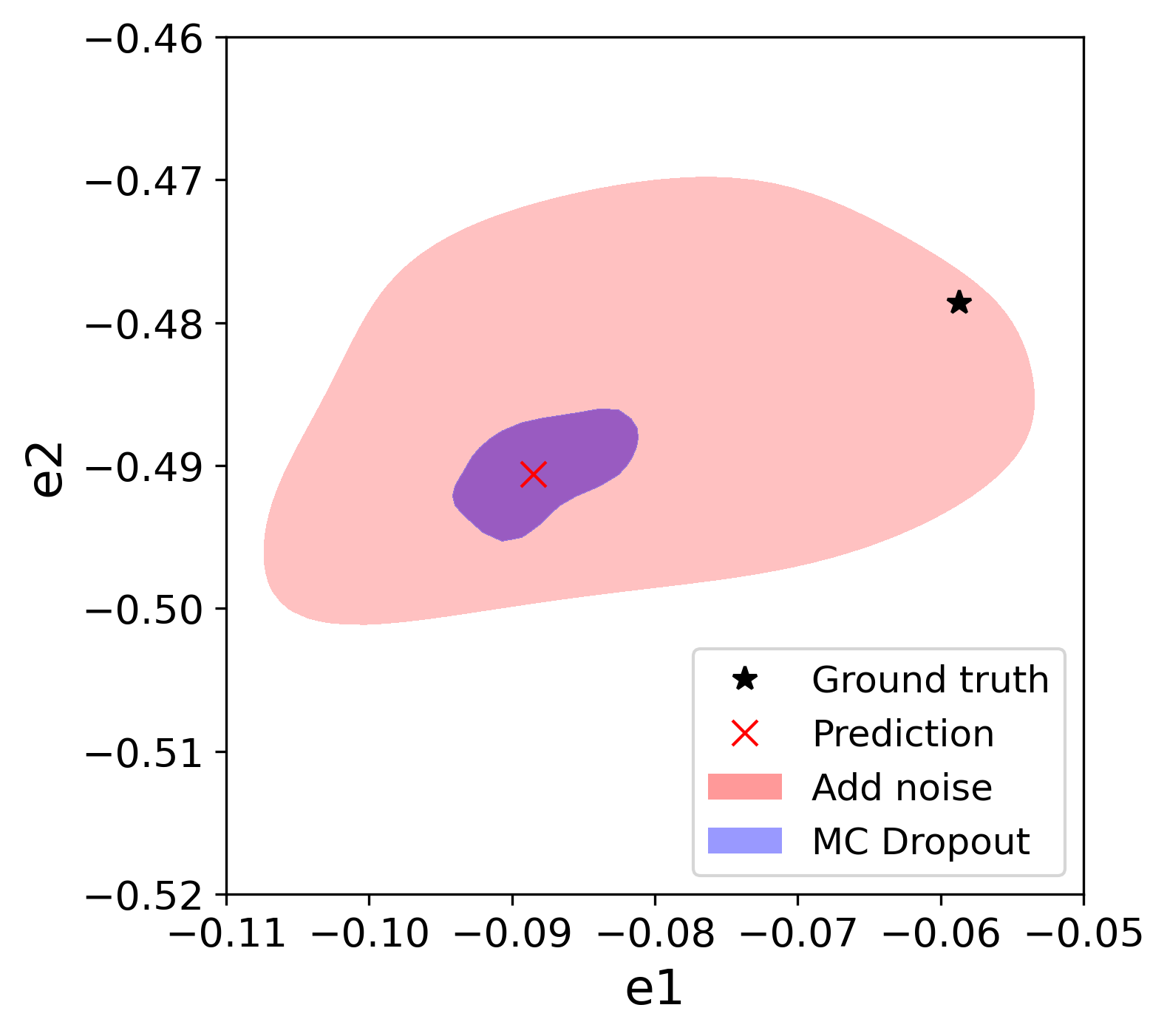}}
    \subfloat[medium SNR]{\includegraphics[width=0.3\linewidth]{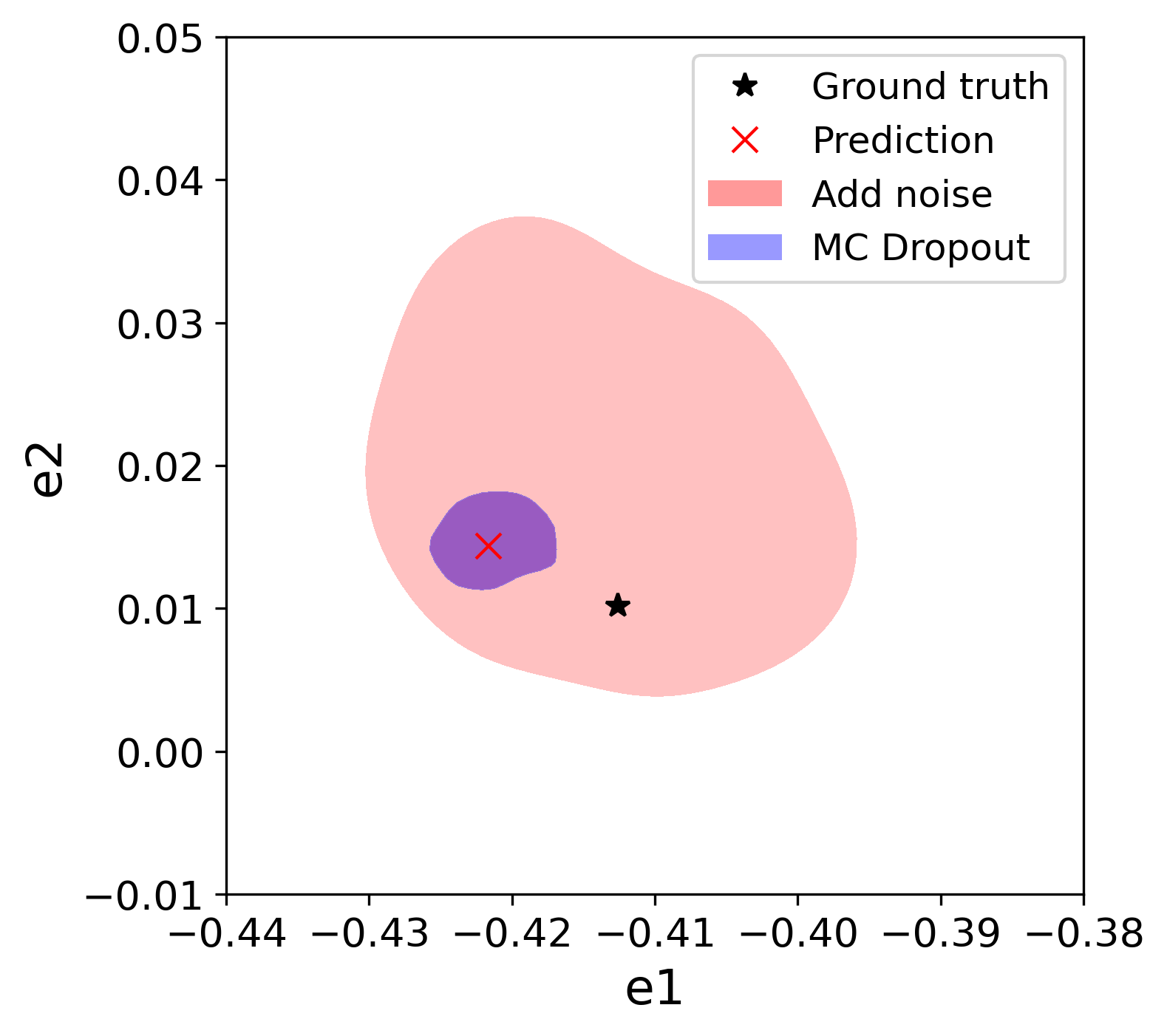}}
    \subfloat[high SNR]{\includegraphics[width=0.3\linewidth]{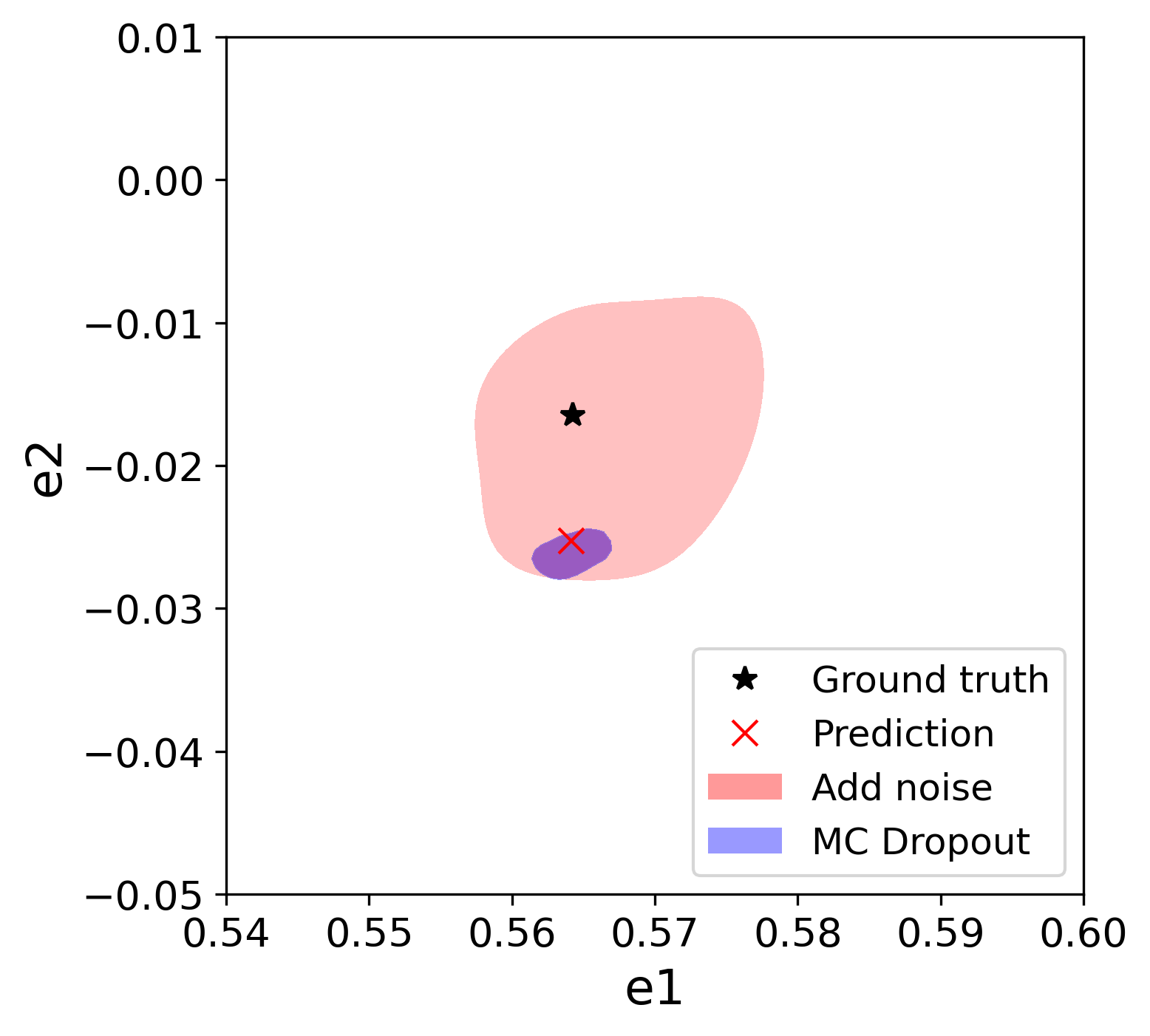}}
    \caption{UQ results for samples with different $M_{00}$.}
\label{fig:uq}
\end{figure}

\subsection{Uncertainty quantification (UQ)}

Like a majority of the deep learning frameworks, DWDN is also a point-prediction method, i.e., the predictions of the network are not associated with error estimates. A large number of UQ approaches are currently applied in the context of AI (artificial intelligence) applications in physics \citep{chen2208}, with associated advantages and disadvantages based on the problem. For instance, sampling over the network weights \citep{neal2012bayesian} or an ensemble-based error estimate \citep{DeepEnsembles2016} may be computationally expensive in our problem, where the individual model training requires 25.5 hours on an Nvidia V100 GPU with 32GB memory. 

In this paper, we perform a two-fold treatment of the uncertainties. The first one aims at estimating the model uncertainties, where we quantify the fluctuations in the predicted deconvolved images due to imperfect model training. Second, we quantify the uncertainty propagated due to noisy input data during the testing phase. Both these treatments are approximations aimed at obtaining a rudimentary understanding of the confidence intervals of our methods, within reasonable computational budgets. In applications involving real observations, several systematic effects may have to be considered in greater detail. 

\subsubsection{Model uncertainty using Monte Carlo methods}
For quantifying model uncertainties, we choose the UQ technique of Monte Carlo (MC) Dropout (\citealt{gal2016dropout}), that provides approximate errors with little computational overhead and minimal changes to the DWDN architecture. Note that the custom losses in Section \ref{sec:custom_loss} are all based on image outputs, and do not consider error estimations.

The implementation of MC Dropout is as follows: First, we train a DWDN without any UQ considerations to obtain an optimized denoising architecture.  Next, we consider an ensemble of $N_{tot}$ for such DWDNs, only differing from each other by a fraction of trained neurons that are re-initialized to a random value. This fraction of weights that are randomized (or `dropped-out') is dictated by the dropout rate $d$. Using the base architecture shown in Figure \ref{fig:framework} with dropout rate $d$, we obtain this ensemble of $N_{tot}$ networks. Since the Deep Neural Networks are over-parametrized, a small number of randomized weights will only result in small fluctuations in the deconvolved output images. 

When a validation image $I$ is forward propagated through each network in the ensemble, they provide individual predictions for the deconvolved images $f^k(I)_{x,y,c}$, where $k=0, 1, \ldots  N_{tot}$. These individual predictions $f^k(I)_{x,y,c}$ are different from each other due to the fact that a different fraction of their network parameters are dropped-out. The mean $\langle f(I) \rangle$ and variance $\sigma_f^2$ of all the individual predictions is calculated as shown in Eq. \ref{eq:mc}. 
\begin{equation}\label{eq:mc}
    \begin{aligned}
    \langle f(I) \rangle &= \frac{1}{N_{tot}} \sum_{k=0}^{N_{tot}} f^k(I)_{x,y,c}, \\
    \sigma_f^2 &= \frac{1}{N_{tot}-1} \sum_{k=0}^{N_{tot}} [f^k(I)_{x,y,c} - \langle f(I) \rangle]^2
    \end{aligned}
\end{equation}

These aggregate mean and variance values will be considered as the UQ predictions from the ensemble. Since the estimates are only computed during the deployment time, there is no change in the network training or associated computational costs. 

\subsubsection{Statistical uncertainty due to noisy image inputs}
In addition to model uncertainty (which corresponds to systematic error in the language of physicists), all the observations contain noise or random uncertainty as well. Since the neural network results (without the MC Dropout at inference) are point estimates, one would ideally process the same image with multiple noise realizations to estimate the noise in any derived quantities. Since this is impossible for real data, we instead generate $N_{tot}$ images with additional noise drawn from the same distribution as the real noise and run those through the network as an example of how to perform uncertainty quantification when only the noisy image is present.  These now form an ensemble from which a noise estimate on the derived quantities can be formed. Since these images now have double the noise variance, strictly speaking, the result is an overestimate of measurement uncertainty.

In the experiment, $N_{tot}=200$ for both approaches and dropout rate $d=0.2$. Results of these two techniques are presented in 
Figure \ref{fig:uq}. We see that the truth is always within the estimated uncertainty region, whose size is likely overestimated. We also see that the size of the statistical uncertainty decreases with increasing SNR, but interestingly the size of the systematic uncertainty also somewhat shrinks with increasing SNR. In other words, as the SNR increases, the network seems to internally produce more consistent results.

We also see that in our case the systematic uncertainty derived using MC Dropout is always significantly smaller than the statistical uncertainty,  even in the right panel of the Figure \ref{fig:uq} which shows the case for one of the highest SNR objects in the test data.  We have inspected a number of high SNR cases and found that the truth almost always lies within the statistical uncertainty errorbar.  In other words, we cannot say with certainty whether the MC Droupout is correctly assessing the systematic error, since that would require a higher SNR object for which the systematic error dominates.

Formally investigating these statistical properties of this uncertainty quantification exceeds the scope of this paper.

\section{Discussion \& Conclusions}

Despite numerous applications of neural networks to astronomical image analysis, the problem of using the information from the known, band dependent and relatively large PSFs has not been addressed directly. While the PSF information is always taken into account in classical image reduction, neural network applications so far usually assume a fixed PSF. This is problematic for several reasons. First, the PSF can undergo weather induced changes in size and shape from observation to observation and from band to band. This means that if a constant PSF is assumed during training, the neural network will have to be retrained for every new set of observations. Similarly, given the large number of possible combinations due to differing band shape and size makes building a grid of models impractical. Therefore, a method is required that is capable of taking the PSF shape as input directly.

The naively obvious approach of simply adding a PSF image to a standard convolutional neural network would not work. Convolutional network inherently assumes that the information is localized in the image coordinate space, while the relation between the PSF image, the ground truth and the observed image is convolutional. We therefore chose a state-of-the art non-blind deconvolution method called Deep Wiener Deconvolution Network. We have trained the network to be able to deconvolve and denoise individual galaxy images. While this approach is not yet ready for prime-time usage, it illustrates a technique that could be used in more complete neural-network based approaches to astronomical image analysis.

Our results can be summarized as follows :
\begin{itemize}
    \item The network is correctly using the information from the provided PSF shape. We checked this by retraining the same model with the mean PSF shape instead and obtained consistently worse results in addition to the expected correlations in the shapes of the output image and the PSF shape. 
    \item   Shape recovery yields the greatest improvement. In that case, the SNR improvement can  be over $50\%$. We also find that while the noise on the individual fluxes is not affected, the color recovery is about $10\%$ better with using actual PSF information.
    \item Without using the PSF shape, there is simply no information in the resulting image to disentangle the effects of the PSF shape from the intrinsic image ellipticity. We have observed that using our best performing network, there is no correlation to be found between the resulting image and the PSF in the current dataset.  However, in the simpler dataset mentioned in \ref{sec:shaperec}, we found residual effects. We applied an alternative network that employs Gaussian regularization, and  it performed nearly as well as the fiducial network on most tests, but completely removed the residual correlations with the PSF. 
\end{itemize}

We have experimented with custom loss functions with some success.  In particular, in addition to the standard pixel-wise $\ell_1$ loss, we have added loss functions that attempt to force the network to improve on the quantities that are of interest to astronomers, by adding loss terms that penalize bad reconstruction of zeroth (total flux), first (astrometry) and second (shape) moments. We have found that naive moment calculation leads to unstable training. Using adaptive moments stabilizes this and the training can proceed with an additional loss contribution of the same order of magnitude.  Unfortunately, we have found that while the loss function improves, the results on the test data do not improve, indicating that we are over training. This could be potentially cured with a larger training dataset, but we leave this investigation for the future. 

Finally, we have investigated methods to understand both the statistical and systematic errors and found that we can correctly quantify the uncertainty in network performance.

The main limitation of the current work is that the network was trained and tested with a single object in each frame. As a result, the network would perform exceptionally well down to unrealistically low SNR (which forced us to impose a SNR cut). The main extension of this work would therefore be to test our approach on images that contain both multiple galaxies in the frame as well as no-galaxies with correct rates and thus use the system as a first step in object detection and deblending. 

\section*{Acknowledgements}
We thank Erin Sheldon for useful discussions. 
This material is partly based upon the work supported by the U.S. Department of Energy, Office of Science, Office of Advanced Scientific Computing Research and Office of High Energy Physics, Scientific Discovery through Advanced Computing (SciDAC) program on ``Accelerating HEP Science: Inference and Machine Learning at Extreme Scales''.
NR's work at Argonne National Laboratory was supported by the U.S. Department of Energy, Office of High Energy Physics. Argonne, a U.S. Department of Energy Office of Science Laboratory, is operated by UChicago Argonne LLC under contract no. DE-AC02-06CH11357.

\bibliographystyle{mnras}
\bibliography{papers}

\end{document}